\documentclass[]{emulateapj-rtx4}
\usepackage{lscape}
\usepackage{longtable}
\usepackage{graphics}
\usepackage{color}

\begin{document}
\title{Infrared and hard X-ray diagnostics of AGN identification\\ 
from the Swift/BAT and AKARI all-sky surveys}

\author{K. Matsuta\altaffilmark{1, 2}, P. Gandhi\altaffilmark{2}, T. Dotani\altaffilmark{2, 1}, 
T. Nakagawa\altaffilmark{2}, N. Isobe\altaffilmark{2},
Y. Ueda\altaffilmark{3}, K. Ichikawa\altaffilmark{3}, \\
Y. Terashima\altaffilmark{4}, 
S. Oyabu\altaffilmark{5}, I. Yamamura\altaffilmark{2, 1}, and \L. Stawarz\altaffilmark{2, 6}}
\affil{
$^1$Department of Space and Astronautical Science, The Graduate University for Advanced Studies,
3-1-1 Yoshinodai, Chuo-ku, Sagamihara, Kanagawa 252-5210, Japan\\
$^2$Institute of Space and Astronautical Science, Japan Aerospace Exploration Agency,
3-1-1 Yoshinodai, Chuo-ku, Sagamihara, Kanagawa 252-5210, Japan\\
$^3$Department of Astronomy, Kyoto University, Kitashirakawa-Oiwake-cho, Sakyo-ku, Kyoto 606-8502, Japan\\
$^4$Department of Physics, Ehime University, 2-5, Bunkyo-cho, Matsuyama, Ehime 790-8577, Japan\\
$^5$Graduate School of Science, Nagoya University, Furo-cho, Chikusa-ku, Nagoya, Aichi 464-8601, Japan\\
$^6$Astronomical Observatory, Jagiellonian University, ul. Orla 171, Krak\'ow 30-244, Poland}

\email{matsuta@astro.isas.jaxa.jp}

\begin{abstract}
We combine data from two all-sky surveys in order to study the connection
between the infrared and hard X-ray ($>$~10~keV) properties for local active galactic nuclei (AGN).
The {\it Swift}/Burst Alert Telescope all-sky survey provides an unbiased, flux-limited selection of hard X-ray detected AGN.
Cross-correlating the 22-month hard X-ray survey with the {\it AKARI} all-sky survey, 
we studied 158 AGN detected by the {\it AKARI} instruments. 
We find a strong correlation for most AGN between the infrared (9, 18, and 90~$\mu$m) and hard X-ray (14--195~keV) luminosities, 
and quantify the correlation for various subsamples of AGN. 
Partial correlation analysis confirms the intrinsic correlation after removing the redshift contribution. 
The correlation for radio galaxies has a slope and normalization identical to that for Seyfert 1s, 
implying similar hard X-ray/infrared emission processes in both. 
In contrast, Compton-thick sources show a large deficit in the hard X-ray band,
because high gas column densities diminish even their hard X-ray luminosities. 
We propose two photometric diagnostics for source classification: 
one is an X-ray luminosity vs. infrared color diagram, 
in which type 1 radio-loud AGN are well isolated from the others in the sample. 
The other uses the X-ray vs. infrared color 
as a useful redshift-independent indicator for identifying Compton-thick AGN.
Importantly, 
Compton-thick AGN and starburst galaxies in composite systems 
can also be differentiated in this plane based upon their hard X-ray fluxes and dust temperatures. 
This diagram may be useful as a new indicator to classify objects 
in new and upcoming surveys such as {\it WISE} and {\it NuSTAR}.
\end{abstract}


\section{Introduction}
Active galactic nuclei (AGN) emit broadband emission over the entire electromagnetic spectrum.
Their typical spectral energy distributions (SEDs) display a strong infrared (IR) bump around 1--100~$\mu$m, 
where a significant portion of their power emerges \citep[e.g.,][]{1993ARA&A..31..473A, 2008NewAR..52..274E}. 
Its source is considered to be a pc-scale torus of dust and gas clouds,
which is heated by radiation from the nucleus and reprocesses this power into the IR, at least for the majority of AGN which are not dominated by beamed emission from relativistic jets \citep[e.g.,][]{2008NewAR..52..274E}. 
The nuclear X-ray emission from AGN is produced by Comptonization of the disk emission 
in a hot corona above an accretion disk \citep{1980A&A....86..121S, 1993ApJ...413..507H, 1998MNRAS.301..915Z}, 
again with a possible exception of jet-dominated sources.

It has been shown that there is a good linear correlation between the logarithms of the observed continuum emission
in the mid-IR and the intrinsic soft X-ray ($< 10$~keV) band luminosities of local AGN,  
which appears to be largely independent of the amount of dust reddening 
\citep[e.g.,][]{2001ApJ...557..626K, 2004A&A...418..465L, 2006A&A...457L..17H, 2008A&A...479..389H, 
2009A&A...502..457G, 2009ApJ...703..390L, 2011arXiv1109.4873A}.
This is not expected according to pioneering AGN dust torus models \citep{1993ApJ...418..673P},
which assumed homogeneous and smooth dust distributions,
because IR emission from the warm-inner clouds is expected to be dramatically scattered and reduced when viewed from an edge-on orientation. 
On the contrary, observations show a tight correlation with little scatter and no IR depletion, 
largely independent of orientation (which ought to correlate approximately with dust reddening) at all AGN luminosities. 
Under the torus paradigm, the correlation may be consistent with models having a clumpy obscuring geometry 
\citep[e.g.,][]{1988ApJ...329..702K, 2005A&A...434..971D, 2006A&A...452..459H, 
2008ApJ...685..160N, 2008A&A...482...67S, 2010A&A...515A..23H, 2010A&A...523A..27H},
in which the inner torus regions can be visible even at edge-on orientations. 

However, previous works are based on small and incomplete samples, 
with selection criteria biased towards bright sources visible in the mid-IR from the ground, 
or sources in which the intrinsic soft X-ray power could be determined. 
Below 10~keV, 
AGN spectra can require detailed modeling due to the presence of several components 
including the underlying AGN power law, 
obscuration, scattering,
non-thermal emission of relativistic jets (in case of radio-loud sources), 
and eventually thermal emission of hot gaseous interstellar medium.
Good quality X-ray spectroscopy is necessary to disentangle these.  
The intrinsic X-ray fluxes of Compton-thick (CT) AGN ($N_{\rm H} >10^{24}$ cm$^{-2}$), especially,   
are fully absorbed in soft X-ray band energies.

In this paper, our aim is to explore the X-ray vs. IR correlation using unbiased samples, 
and then define new and simple photometric diagnostics for AGN classification. 
We use the {\it Swift}/Burst Alert Telescope (BAT) hard X-ray ($>10$~keV) survey as the base AGN selection catalog. 
The hard X-ray energy band is rather insensitive to the intervening column density up to mildly CT columns. 
The survey reflects the intrinsic luminosity of the source and provides samples largely unbiased by obscuration. 
We use the {\it Swift}/BAT 22-month Source Catalog herein \citep{2010ApJS..186..378T}. 
Several recent works explored the connection of hard X-rays with near-IR powers and with mid-IR emission line properties 
\citep[e.g.,][]{2008ApJ...684L..65M, 2009ApJ...698..623D, 2009ApJ...700.1878R, 2010ApJ...716.1151W},
but correlation with an all-sky mid-IR survey at high angular resolution is still lacking. 

High angular resolution is crucial in order to properly separate AGN from stellar emission in the IR 
\citep[e.g.,][]{2009A&A...502..457G, 2009A&A...495..137H, 2010MNRAS.402.1081V, 2011MNRAS.414.1082M}.
For many years, {\it IRAS} provided the most complete all-sky survey \citep[][]{1994yCat.2125....0J}, 
but with an angular resolution only of order arcmins. 
The {\it AKARI} satellite \citep{2007PASJ...59S.369M} has now completed its all-sky survey which is several times more sensitive than {\it IRAS}, 
and at a much higher angular resolution of order arcsecs,
(7\arcsec\ and 48\arcsec\ for the mid- and far-IR instruments, respectively). 
The {\it AKARI} survey thus provides the best IR sample for cross-correlation with {\it Swift} AGN. 
We use data available in the {\it AKARI} Point Source Catalogs \citep[{\it AKARI}/PSC;][]{2010A&A...514A...1I, 2010yCat.2298....0Y} 
over a range of wavelengths from 9--160~$\mu$m. 
Although {\it AKARI}'s angular resolution is lower than large ground telescopes 
(e.g., the diffraction limit of the Very Large Telescope/VISIR $\sim$ 0.3\arcsec\ in the {\it N}-band),
our study has the capability of probing predominantly nuclear emission in the mid-IR (in addition to hard X-rays). 
This is because torus emission typically dominates over nuclear stellar emission in AGN host galaxies 
in the wavelength regime of 10--20~$\mu$m \citep[e.g.,][]{2009ApJ...705..298M}.

The completeness of the base samples (largely unbiased by obscuration) allows us to examine and design various color correlation plots using the hard X-ray and IR powers 
which can be useful for source classification in blind surveys. 
Such diagnostics can be particularly powerful if they are redshift independent and can be shown to hold 
when redshift {\it K}-correction is accounted for. 
We propose one such color--color diagnostic, 
and show that it also provides the capability of distinguishing starbursts from CT AGN. 
This has been an important problem for many years because star formation and obscured accretion activity often occur together in composite systems 
\citep[cf. the well known \lq starburst--AGN connection\rq; e.g.,][]{1998ApJ...505..174G, 1998ApJ...498..579G, 2006ApJ...649...79S,
1996ARA&A..34..749S, 1998ApJ...498..579G, 1998ApJS..117...25M, 2007ApJS..171...72I}. 
The observed soft X-ray and IR appearance is not easily distinguishable in these cases. 
Our hard X-ray selection now allows us to do this. 
The simple photometric diagnostics can be useful tools for new and upcoming hard X-ray surveys with {\it NuSTAR}, 
{\it ASTRO-H} and in the IR with {\it WISE}. 

This paper is organized as follows. 
In section 2, we present details of the sample selection and catalog cross-matching. 
Section 3 shows the main results regarding the hard X-ray and IR luminosity correlation and its statistics. 
Section 4 discusses the implications of the results and provides new diagnostic plots for AGN classification. 
Section 5 lists the main conclusions. This work is closely related to that of Ichikawa et al. (2012),
in which the relation between the IR/X-ray spectra of Seyferts and  new type AGN are studied 
by using the {\it Swift}/BAT 9-month AGN Catalog \citep{2008ApJ...681..113T} 
and the {\it AKARI}/PSC complemented by those with {\it IRAS} and {\it WISE} to ensure high completeness of cross identification.
The larger 22-month BAT sample used as the starting catalog for the present work includes several heavily CT AGN and radio-loud (RL) sources, 
and we are able to compare the regions occupied by these various AGN classes in the luminosity correlation plane. 
We also provide detailed statistical tests and correlation fits, and construct AGN photometric classification plots.


A flat Universe with a Hubble constant $H_0 = 71$~km~s$^{-1}$~Mpc$^{-1}$,
$\Omega_\Lambda = 0.73$ and $\Omega_{\rm M} = 0.27$ is assumed
throughout this paper.  

\section{Data Selection}
We used the following two all-sky survey catalogs to study the correlation in a complete, 
flux-limited sample and examine the luminosity correlation between the observed IR and hard X-ray bands of local AGN ($z<0.1$); 
the {\it Swift}/BAT 22-month Source Catalog 
and the {\it AKARI}/PSC. 

\subsection{The {\it Swift}/BAT 22-month Source Catalog}
The prime objective of the {\it Swift}/BAT is the study of gamma-ray bursts,
covering the hard X-ray band of 14--195~keV.
In the 22-month BAT catalog, the total source count is 461 (above a significance of 4.8$\sigma$), 
of which 268 (58\%) are AGN.
The catalog includes source information of optical counterpart position, 
X-ray flux, flux error, luminosity, redshift, and AGN type.
We adopted the position of the optical counterpart as the source position.
The counterpart position error is usually less than 4\arcsec .    
The detection limit of the catalog is $2.3\times 10^{-11}$~erg cm$^{-2}$s$^{-1}$ (for 50\% sky coverage).
The faintest source included in the catalog has $F_{\rm 14-195\ keV}=0.28\times 10^{-11}$~erg cm$^{-2}$s$^{-1}$. 
Although progressively-deeper BAT catalogs are becoming available \citep[e.g.,][]{2010A&A...524A..64C}, 
the flux limit of the {\it AKARI} all sky survey is well-matched for detection of a large fraction of the 22-month X-ray catalog in the infrared, 
as we show below. 
Thus, we have restricted ourselves to the 22-month sample for AGN analysis in the present work. 
We refer to deeper samples only with regard to X-ray emission from starbursts in Section~\ref{discuss:ratio}.

The {\it Swift}/BAT survey produces a long-term average flux, 
so variability is expected to be partially smoothed over,
as compared to large stochastic fluctuations on weekly or monthly timescales. 
We defined a minimum flux error of 10\%, 
in order to account for low-order long term variations.

We adopted the spectroscopic AGN types from the BAT catalog for most sources, 
but for some AGN we checked other works because these source classifications are ambiguous
[NGC~235A,  NGC~612, ESO~549-G049, NGC~3998, WKK~1263, 4C~50.55, and 3C~445, \citet{2010A&A...518A..10V};
NGC~4945, \citet{1996A&A...308L...1M};
NGC~4992, \citet{2007A&A...467..585B}; 
Circinus Galaxy, \citet{2000MNRAS.318..173M};
3C~345, \citet{1997ApJ...480..596U}].
We excluded AGN defined as `confused and confusing with nearby sources' in the BAT catalog.  
Finally, we identified all radio-loud AGN in the combined sample, 
including type 1 objects such as broad-line radio galaxies and blazars correctly recognized in the BAT classification, 
and also type 2 sources including narrow-line radio galaxies and low-power FR I radio galaxies which in the BAT catalog are classified simply 
as Seyfert galaxies (or LINERs) based solely on the characteristics of their optical spectra and regardless of their radio properties.

\subsection{The {\it AKARI} Point Source Catalogs}\label{AKARI_catalog}
{\it AKARI} has two instruments; the Infrared Camera (IRC) and the Far-infrared Surveyor (FIS).
The IRC band centers are 9 and 18~$\mu$m, and the FIS band centers are 65, 90, 140, and 160~$\mu$m.
The {\it AKARI}/PSC contains the positions and fluxes of 870,973 (IRC) and 427,071 (FIS) sources. 
In the {\it AKARI}/PSC, the IRC 80\% flux completeness limits are 0.12 and 0.22~Jy, respectively.
Also, the FIS 80\% completeness corresponds to 3.3, 0.43, 3.6, and 8.2~Jy, respectively.
The FIS may be incomplete or show a larger flux error 
due to detector saturation for sources brighter than $\sim$ 100~Jy.
The angular resolution is 7\arcsec\ (IRC) and 48\arcsec\ (FIS), respectively.

In the {\it AKARI}/PSC, there is a quality flag, {\it FQUAL}, par band.
This flag is a four-level (3-0) flux quality indicator, 
and the {\it AKARI} team recommends using only sources with {\it FQUAL} = 3 for secure scientific analysis. 
Although sources with {\it FQUAL} = 2 or 1 may have flux values reported \citep{2010yCat.2298....0Y}, 
we used only sources that are confirmed and described by the quality flag {\it FQUAL}= 3, 
indicating a high reliability of detection and flux accuracy.  

\subsection{Source identification}\label{sec:sourceid}
We searched for IR counterparts within a 10\arcsec\ radius for the IRC, 
and 20\arcsec\ radius for the FIS around the position of every optical counterpart 
listed in the BAT catalog using the {\it AKARI} Catalog Archive Server \cite[CAS;][]{2011PASP..123..852Y},
which provides user interfaces to search and obtain {\it AKARI}/PSC information. 
We adopted these search radii based upon the $\sim 3\sigma$ position uncertainties for each of the {\it AKARI}/PSC. 

We concentrated on  three {\it AKARI} IR bands (9, 18, and 90~$\mu$m) throughout this paper
because other FIS bands have poorer sensitivities. 
We summarized the selected source parameters (name, IR fluxes, X-ray luminosity, redshift, AGN type)
in Table \ref{tb:CASresult-detail}.
Additionally, we summarized the total sample statistics for the three {\it AKARI} bands in Table~\ref{tb:CASresult}.
For the 268 AGN included in the BAT catalog,
158 AGN ($\sim$ 59\%) are detected in at least one of the IRC and FIS bands
(111 in 9~$\mu$m; 129 in 18~$\mu$m; 113 in 90~$\mu$m).
These 158 AGN are classified into
81 Seyfert 1 (Sy1) type AGN ($\sim$ 51\%, including Seyfert 1, 1.2, and 1.5),
67 Seyfert 2 (Sy2) type AGN  ($\sim$ 42\%, including Seyfert 1.8, 1.9, and 2),
4 Low-ionization nuclear emission-line regions (LINERs; $\sim 3\%$),
and 6 blazars ($\sim$ 4\%).
Four blazars are flat-spectrum radio quasars (FSRQs),
and two are BL Lac objects.

For the purpose of investigating the IR vs. hard X-ray luminosity correlation for various classes of objects, 
we divide our sample into the following ten categories:

1. ``All''

2. ``All, ex CT''

3. ``Sy1''

4. ``Sy1, ex RL''

5. ``Sy2''

6. ``Sy2, ex CT''

7. ``CT''

8. ``RL''

9. ``Blazar''

10. ``RL + Blazar''.

``All'' contains all sources detected by the {\it AKARI} and {\it Swift}/BAT.
``All, ex CT'' excludes CT AGN from ``All''.
``CT'' is defined as sources obscured by a gas column density 
$N_{\rm H}\ge 1.5\times10^{24}$~cm$^{-2}$ as measured from X-ray spectroscopy.  
``Sy1'' combines type 1 AGN including radio-quiet objects (Seyferts) 
and radio-loud sources such as broad-line radio galaxies (BLRGs) and quasars.
``Sy1, ex RL'' excludes RL AGN from ``Sy1''.
``Sy2'' combines type 2 AGN including CT objects, radio quiet Seyferts, and radio-loud sources 
such as narrow-line radio galaxies and low-power FR I radio galaxies.
``Sy2, ex CT'' excludes CT AGN from ``Sy2''.
``RL'' combines all radio-loud AGN in the sample except of blazars, i.e., all radio galaxies of different morphological and spectral types.
``Blazar'' combines FSRQs and BL Lacs. Finally, 
``RL + Blazar'' merges the two radio loud subsamples.

The criterion of classification as a RL (non-blazar) AGN is that 
the radio luminosity density at 5~GHz is at least $10^{32}$~erg~s$^{-1}$~Hz$^{-1}$ and 
the radio to optical {\it B}-band flux density ratio ($R_B$) satisfies log $R_B \ge$  1
 \citep{1989AJ.....98.1195K, 1999AJ....118.1169X, 2008MNRAS.390.1217M}.   
RL objects are identified in the final column of Table \ref{tb:CASresult-detail}. 
This list includes 3C and 4C sources from the Cambridge catalogs of radio galaxies, 
as well as other famous RL sources such as Cen A. 

A few additional objects were also identified as RL based on a detailed inspection of the available data and literature.
NGC~1052 is a Gigahertz-peaked spectrum (GPS) source, with a spectral peak at 10~GHz \citep{2003A&A...401..113V}. 
Although the 5~GHz power is lower than the RL classification threshold,
this source displays strong parsec-scale radio emission probably related to newly born jets, and is classified as RL. 
[HB89]1821+643 is a peculiar radio source, 
lying at the boundary of the radio-quiet/radio-loud divide \citep{2001ApJ...562L...5B}. 
The source is a highly luminous quasar in X-rays 
\citep[$L_{\rm X}>1\times10^{47}$ erg s$^{-1}$;][]{{2010MNRAS.402.1561R}}
and lies at the center of a massive galaxy cluster.
It shows an FR I like radio morphology \citep{2001ApJ...562L...5B}. 
As we discuss below, the {\it AKARI} observations now show 
that the IR properties of [HB89]1821+643 are consistent with other RL sources, 
so we include it within the RL classification herein, as a radio-intermediate (RIM) object.
Similarly, two other RIM sources are  
[HB89]0241+622 \citep{2008MNRAS.390.1217M} and Mrk~1501 \citep{2005ApJ...618..108B}.

There are several narrow-line Seyfert 1s (NLS1s) in the sample, 
but these are not distinguishable from the other Seyferts so we do not treat them separately.

\section{Results}


\subsection{Redshift, Flux distribution and Completeness}\label{redshift}
\begin{figure}[!ht]
 \begin{minipage}{0.5\hsize}
  \begin{center}
   \includegraphics[width=65mm]{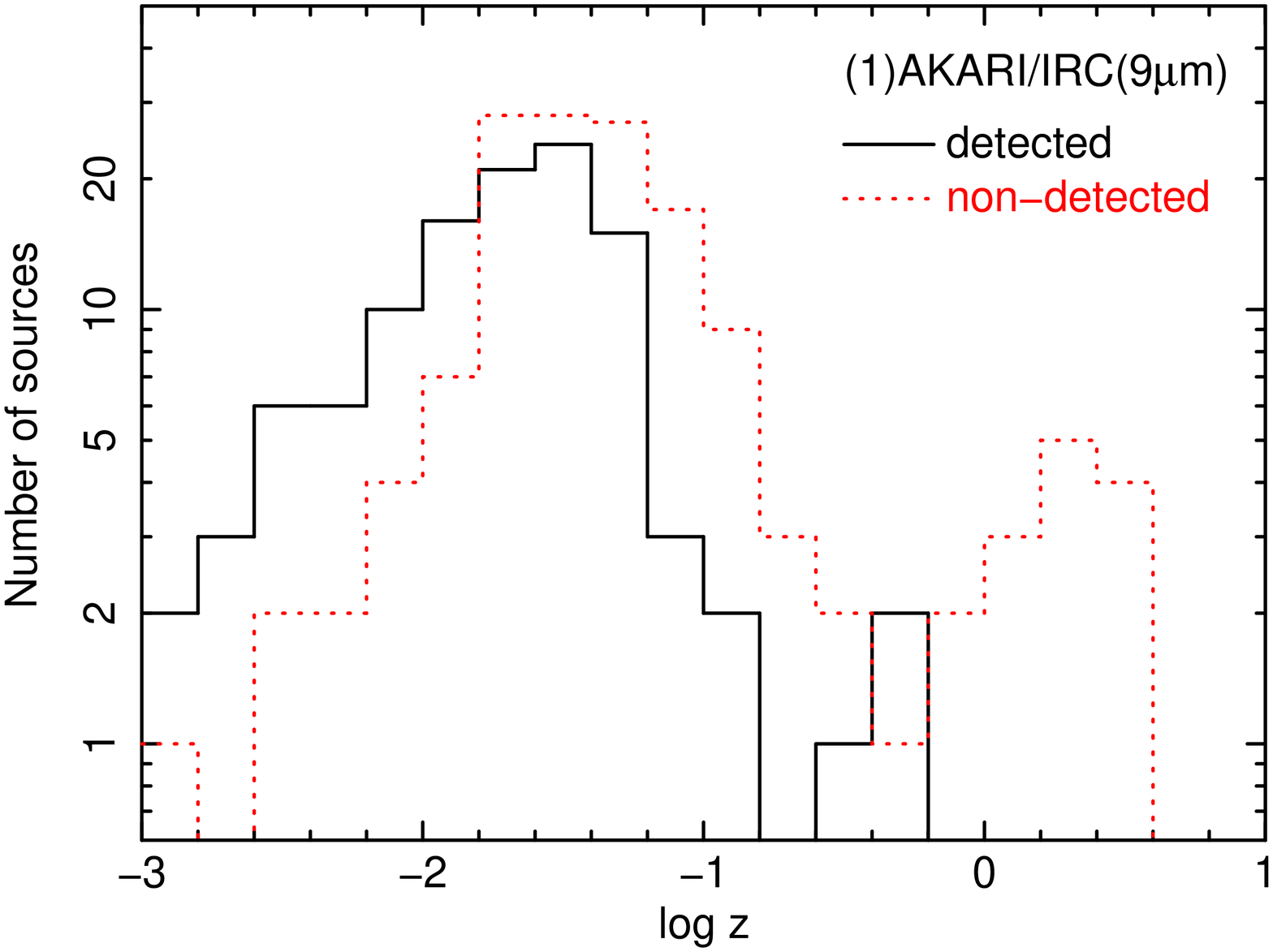}
  \end{center}
 \end{minipage}
 \begin{minipage}{0.5\hsize}
  \begin{center}
   \includegraphics[width=65mm]{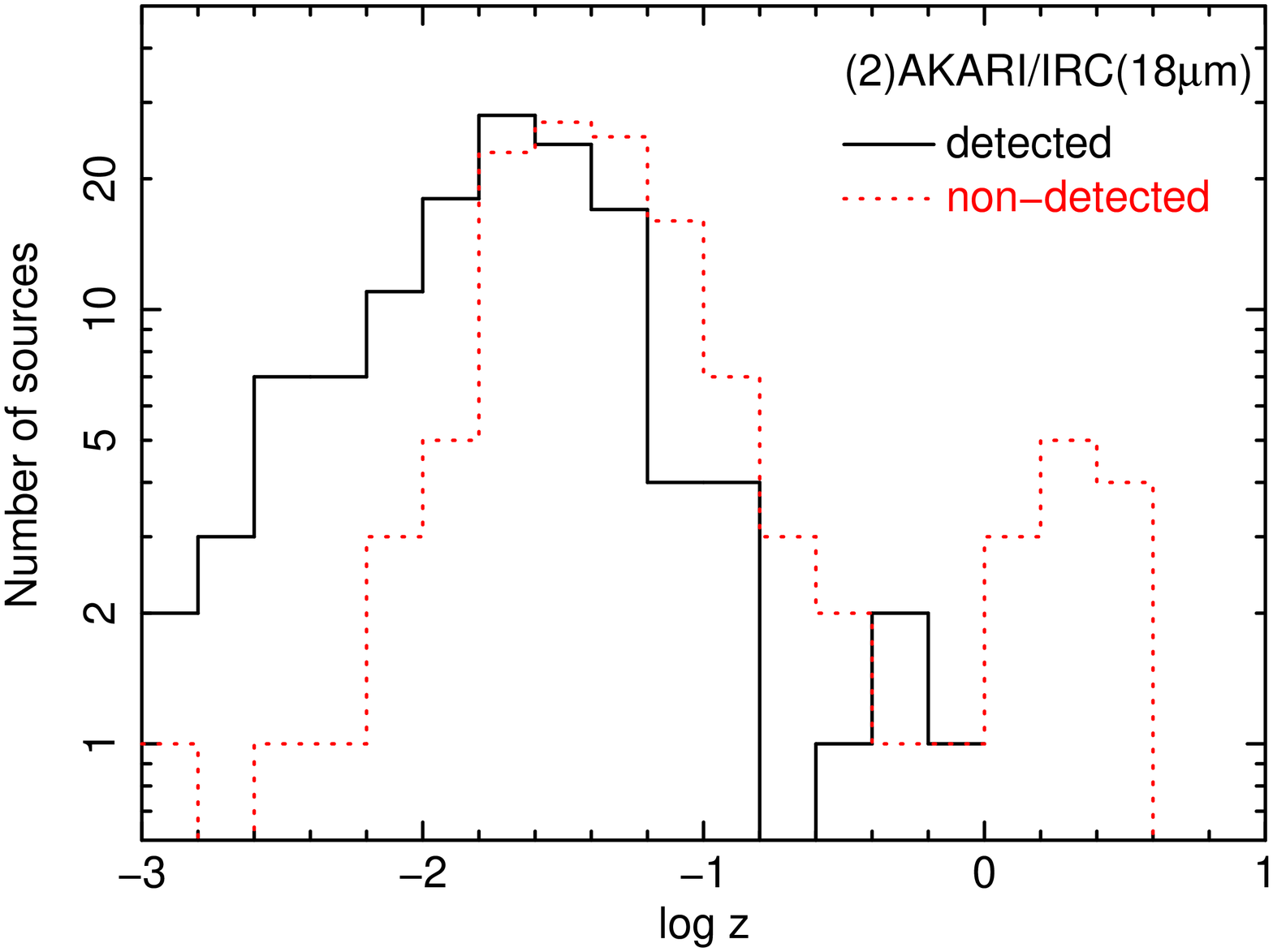}
  \end{center}
 \end{minipage}
 \begin{minipage}{0.5\hsize}
\vspace{5pt}
  \begin{center}
   \includegraphics[width=65mm]{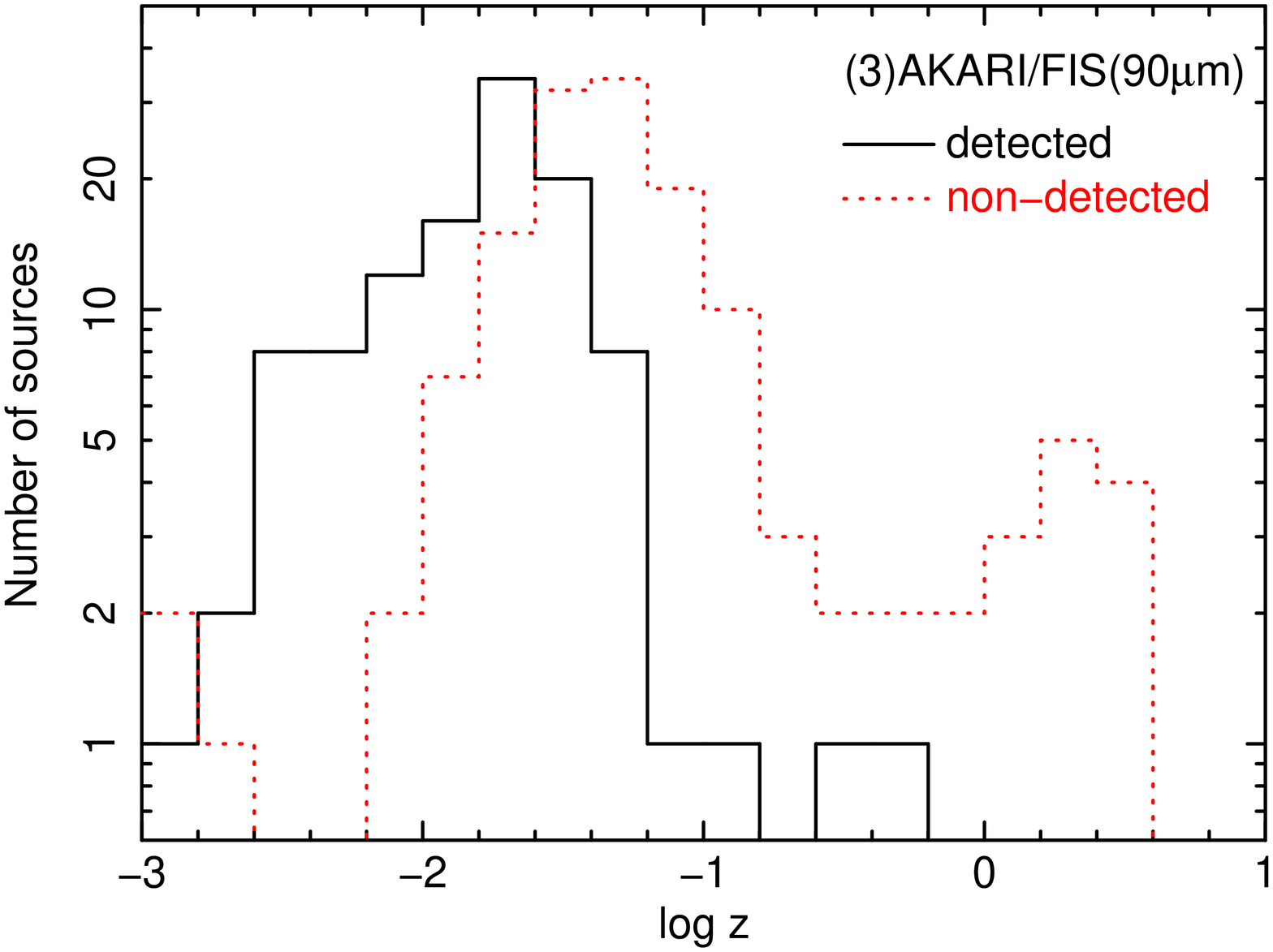}
  \end{center}
 \end{minipage}
\caption{Redshift distribution of all AGN detected in the {\it Swift}/BAT.
The values of redshift were taken from those listed in the 22-month {\it Swift}/BAT hard X-ray survey catalogue 
\citep{2010ApJS..186..378T}.
The abscissas are logarithmic redshift.
The ordinates are total number of sources.
Black solid lines show the {\it AKARI} detected sources ({\it AKARI}/PSC; {\it FQUAL} = 3 only),
while red dotted lines show the sources not listed in the {\it AKARI}/PSC.}
\label{fig:redshift}
\end{figure}

Figure~\ref{fig:redshift} shows the redshift distribution of the {\it Swift}/BAT AGN, 
split into the classes of {\it AKARI}-detected or non-detected objects. 
The median redshift of the detected sources ($z_{\rm med}$) is $\approx 0.0222$,
so most are local AGN.
There are only 8/158 sources (5\%) with $z >0.1$.
The detected source with the maximum redshift is 3C~454.3 ($z_{\rm max}\approx$ 0.8590).
The low-{\it z} population is dominated by ``regular" (radio-quiet) Seyferts, 
while the high-{\it z} population is made mostly from radio-loud AGN (blazars, luminous radio galaxies).
The $z_{\rm med}$ for those not detected in any band is $z_{\rm med}\approx 0.0517$.

\begin{figure}[!ht]
 \begin{minipage}{0.5\hsize}
  \begin{center}
   \includegraphics[width=65mm]{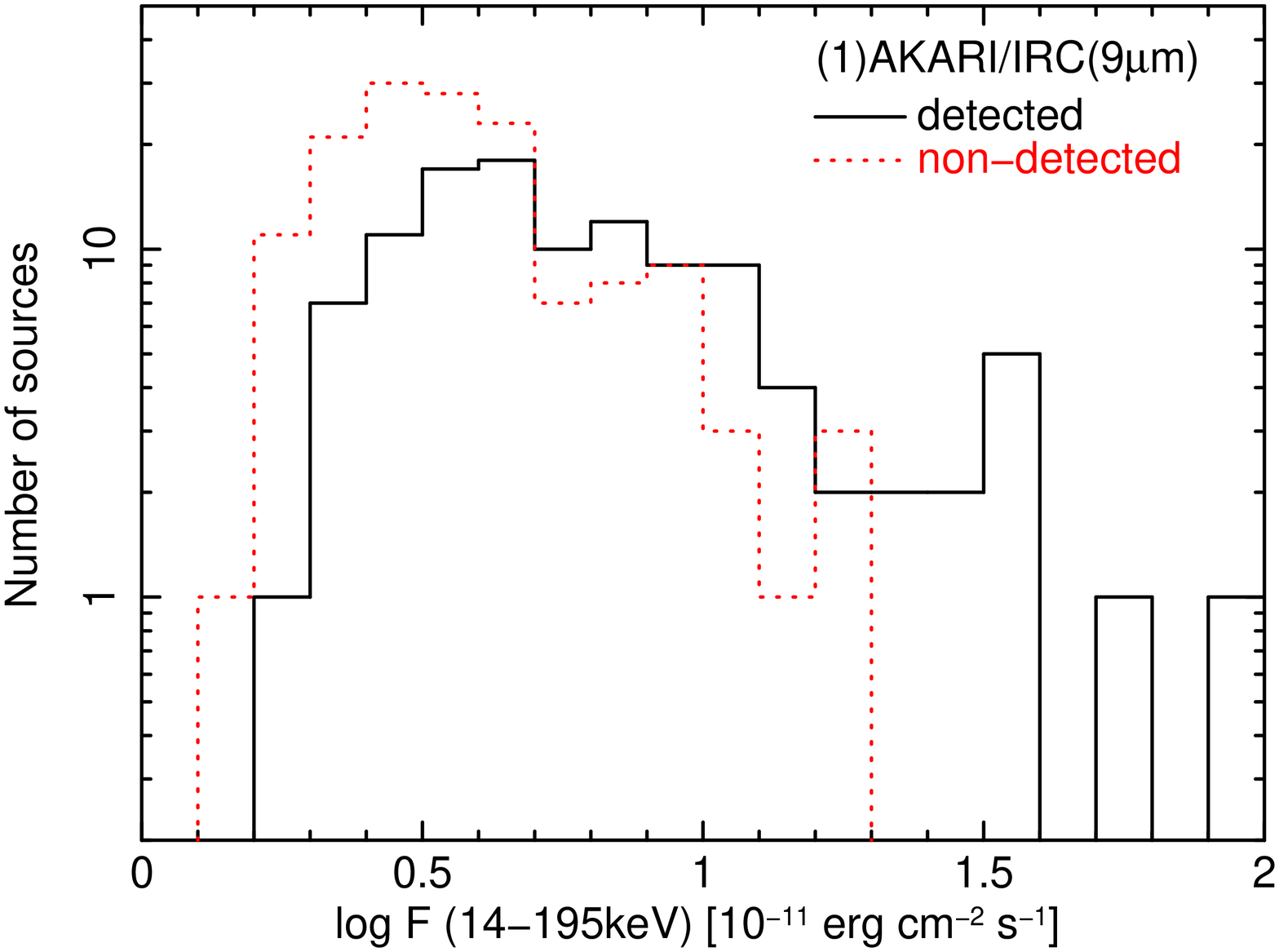}
  \end{center}
 \end{minipage}
 \begin{minipage}{0.5\hsize}
  \begin{center}
   \includegraphics[width=65mm]{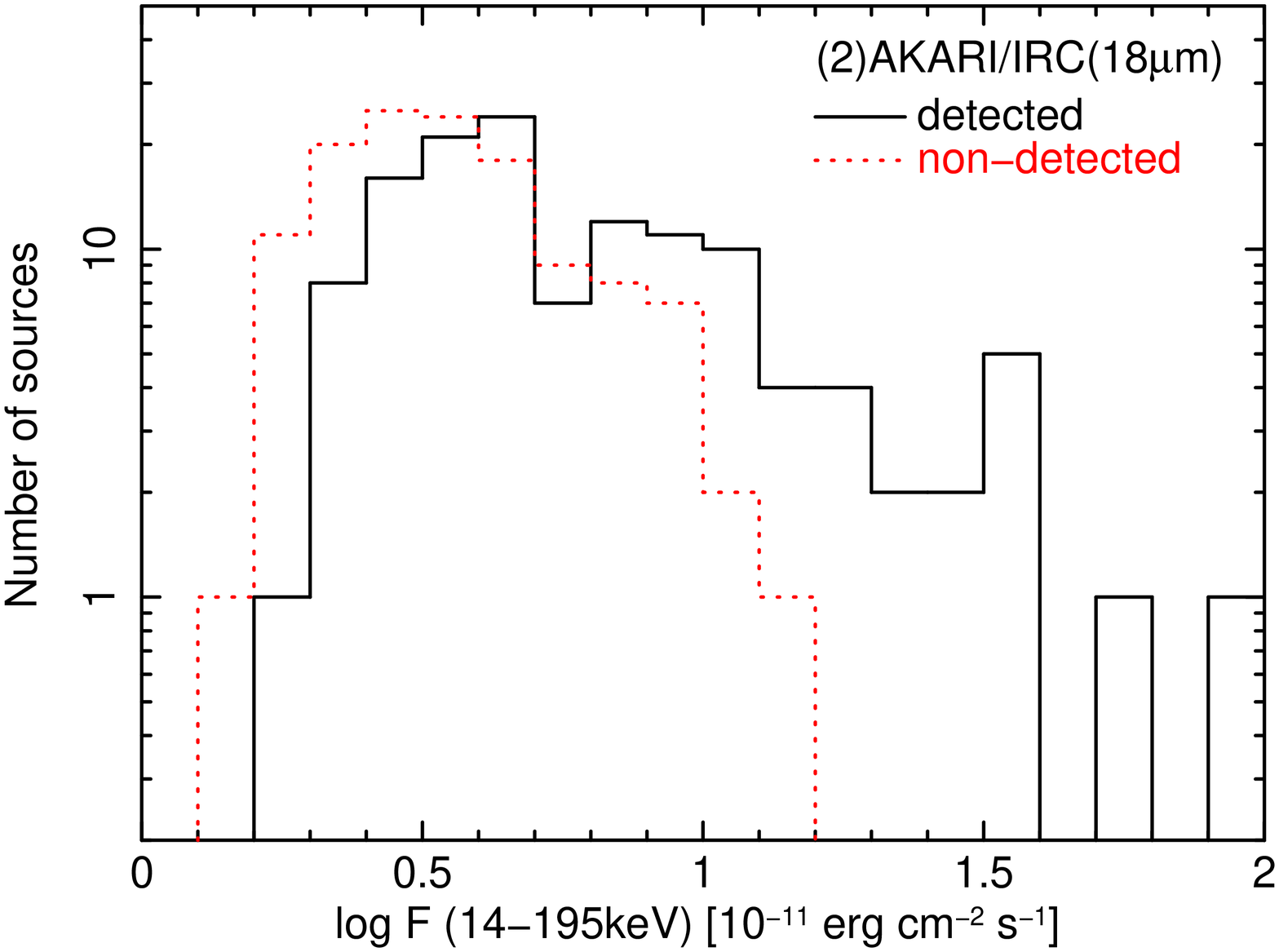}
  \end{center}
 \end{minipage}
 \begin{minipage}{0.5\hsize}
\vspace{5pt}
  \begin{center}
   \includegraphics[width=65mm]{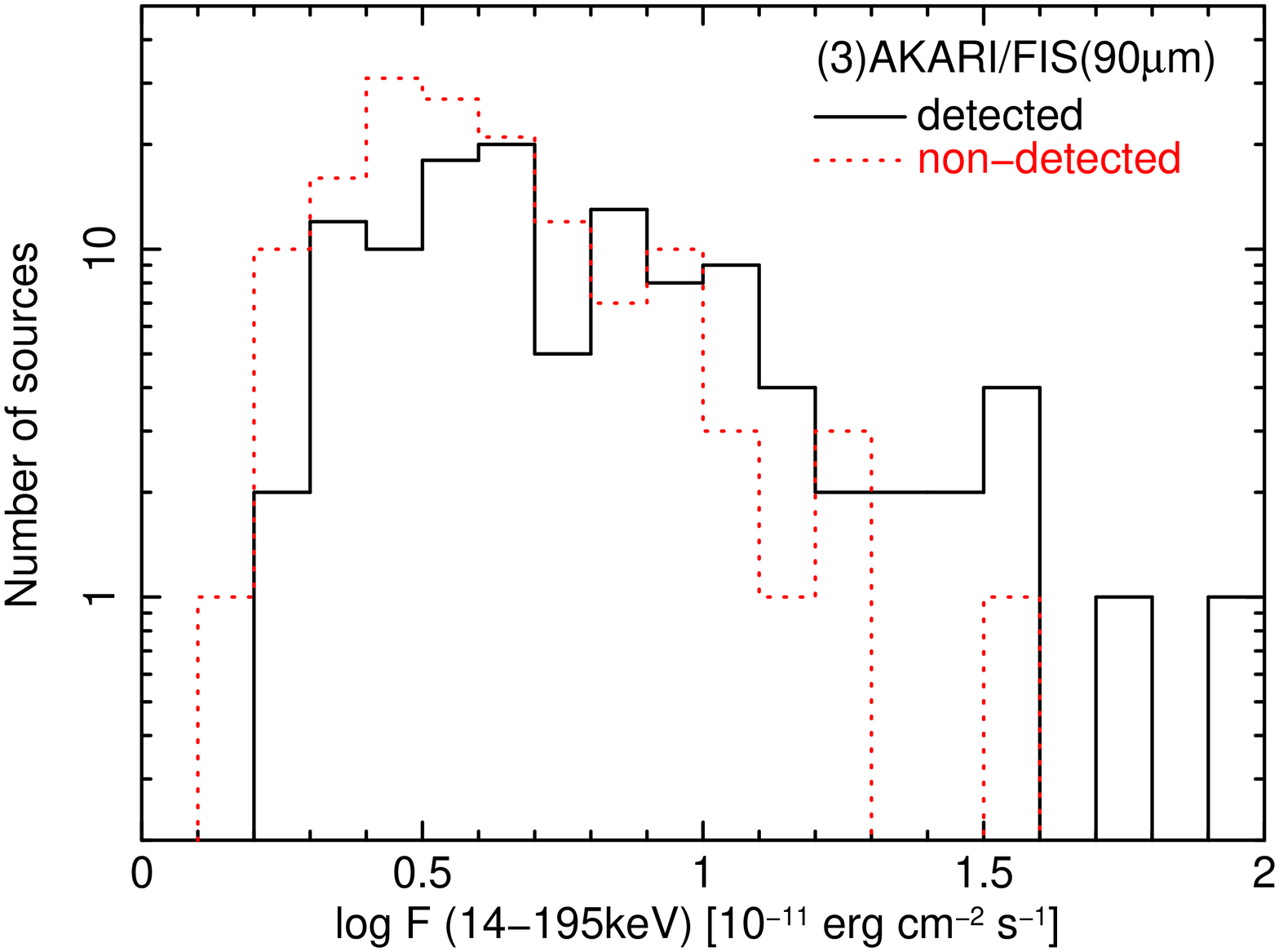}
  \end{center}
 \end{minipage}
\caption{Hard X-ray flux distributions for the {\it Swift}/BAT sources.
The abscissa is logarithmic hard X-ray flux.
The ordinate is the number of sources.
Black solid lines show the {\it AKARI} detected sources,
while red dotted lines show those are not detected by {\it AKARI} in 3$\sigma$.}
\label{fig:Hist-Fx}
\end{figure}

Figure~\ref{fig:Hist-Fx} shows the hard X-ray flux distribution of the {\it Swift}/BAT AGN.
The X-ray fluxes of the {\it AKARI} detected sources are higher than the non-detected sources on average.
The median X-ray fluxes of the {\it AKARI} detected/non-detected sources are
5.07/3.38, 4.70/3.36, and $4.70/3.45~\times~10^{-11}$~erg cm$^{-2}$ s$^{-1}$
in 9, 18, and 90~$\mu$m, respectively. 

Our cross-correlation between the all-sky X-ray and IR samples is about 60\% complete, 
with the X-ray sources non-detected in the IR being fainter than the detected ones, on average. 
It has been shown by Ichikawa et al. (2012) 
that supplementing the {\it AKARI} data with the deeper {\it WISE} samples maintains the luminosity correlation. 
Overall, none of the statistical inferences presented in the sections below should be affected by incompleteness of the cross-matching. 


\subsection{Hard X-ray to IR luminosity correlation}\label{LLpot}
Figure~\ref{fig:LLplot} shows the luminosity correlation  for the detected sources between hard X-ray (14--195~keV)
and IR (9, 18, and 90~$\mu$m) bands. 
We find that there is a strong linear correlation for most AGN between the logarithms of the observed
IR and hard X-ray luminosities over four orders of magnitude.
Here, we calculated the absolute luminosities by using the redshift listed in the BAT catalog \citep{2010ApJS..186..378T}.
In principle, {\it K}-correction may become important for high-{\it z} objects, e.g., blazars.
Our samples are mainly local AGN, 
but blazars ({\it z }= 0.0533--0.8590) are also included. 
If we apply {\it K}-correction to AGN at $z\sim$ 0.9,
the maximum effect for $L_{\rm X}$ is $\sim$ 10\% assuming a power law spectrum with a photon index of $\Gamma \sim 1.9$, 
which is a typical value for the majority of AGN in our sample. 
This is a small effect especially when comparing logarithmic luminosities, so we did not apply any {\it K}-corrections.
 
We tested the luminosity correlation by using the algorithm of \cite{1990ApJ...364..104I}, 
which is used when the nature of the scatter of points in a correlation is ill understood. 
We computed the linear regression coefficients by one of the methods of the algorithm, ``OLS bisector'',
recommended if the goal is to determine the functional relation between the two axes. 
The formula is given by
\begin{equation}
\log L_{\rm IR} = a + b \log L_{\rm X},
\label{eq:OLS}
\end{equation}
where $L_{\rm IR}$ and $L_{\rm X}$ are observed luminosities in IR (9, 18, and 90~$\mu$m) and hard X-ray (14--195~keV).
The variables {\it a} and {\it b} are the intercept and slope of the fitting result, respectively.
At 9~$\mu$m, 
the luminosity correlation of the subsample ``Sy1'' is described as 
\begin{eqnarray}
\nonumber\log\Bigl(\frac{\lambda L_{\rm 9\mu m}}{10^{43}~{\rm erg~s^{-1}}}\Bigr) &=(0.15\pm0.05)& \\
&\hspace{-40pt}+(0.94\pm0.06)&\hspace{-20pt}\log\Bigl(\frac{L_{\rm X}}{10^{43}~{\rm erg~s^{-1}}}\Bigl).
\end{eqnarray}
The fit parameters for other bands are listed in Table~\ref{tb:LLfit}, 
and are plotted  in Figure~\ref{fig:LLplot} along with the 1$\sigma$ and 2$\sigma$ uncertainties on the correlation normalization. 
Also, we summarize the fitting results with respect to various subsamples in Table~\ref{tb:LLfit}.

We checked two correlation coefficients for every subsample, 
the Spearman's Rank correlation coefficient ($\rho$) and the partial correlation coefficient ($\rho_{\cdot z}$).
The $\rho$ and $\rho_{\cdot z}$ correlation tests 
return values in the interval [-1.0, 1.0].
The $\rho_{\cdot z}$ statistic is especially important 
because artificial correlations between luminosities may be induced in a flux limited-sample even in the absence of any intrinsic correlation, 
but the partial correlation coefficient can account for this by excluding the effect of redshift. 
A large value implies a significant positive, linear correlation.   
These test coefficients values are also listed in Table~\ref{tb:LLfit}.

``Blazar'' and ``RL + Blazar'' typically show very high values of $\rho$ and $\rho_{\cdot z}$ 
in all three {\it AKARI} bands. 
But only a handful of objects included in both subsamples are among the most luminous and distant sources in the entire sample considered, 
so the correlation results for these may be biased by small number statistics and flux limits.
The $\rho$ and $\rho_{\cdot z}=1$ values in these cases must be treated with caution. 
The other subsamples with the highest values of $\rho$ and $\rho_{\cdot z}$ are the ``Sy1'' and ``All, ex CT'' subsamples. 
``Sy2'' also show a correlation, if CT sources are excluded, 
though somewhat weaker than ``Sy1''. 
CT sources will be discussed separately in section~\ref{discuss:CT-AGN}. 
The correlation coefficients generally decrease towards longer IR wavelengths. 

\section{discussion}\label{sec:discus}
\subsection{Mid-infrared to hard X-ray correlation\\ for Seyfert AGN}\label{discuss:LL-all}
\begin{figure}[!ht]
 \begin{minipage}{0.33\hsize}
  \begin{center}
   \includegraphics[width=57mm]{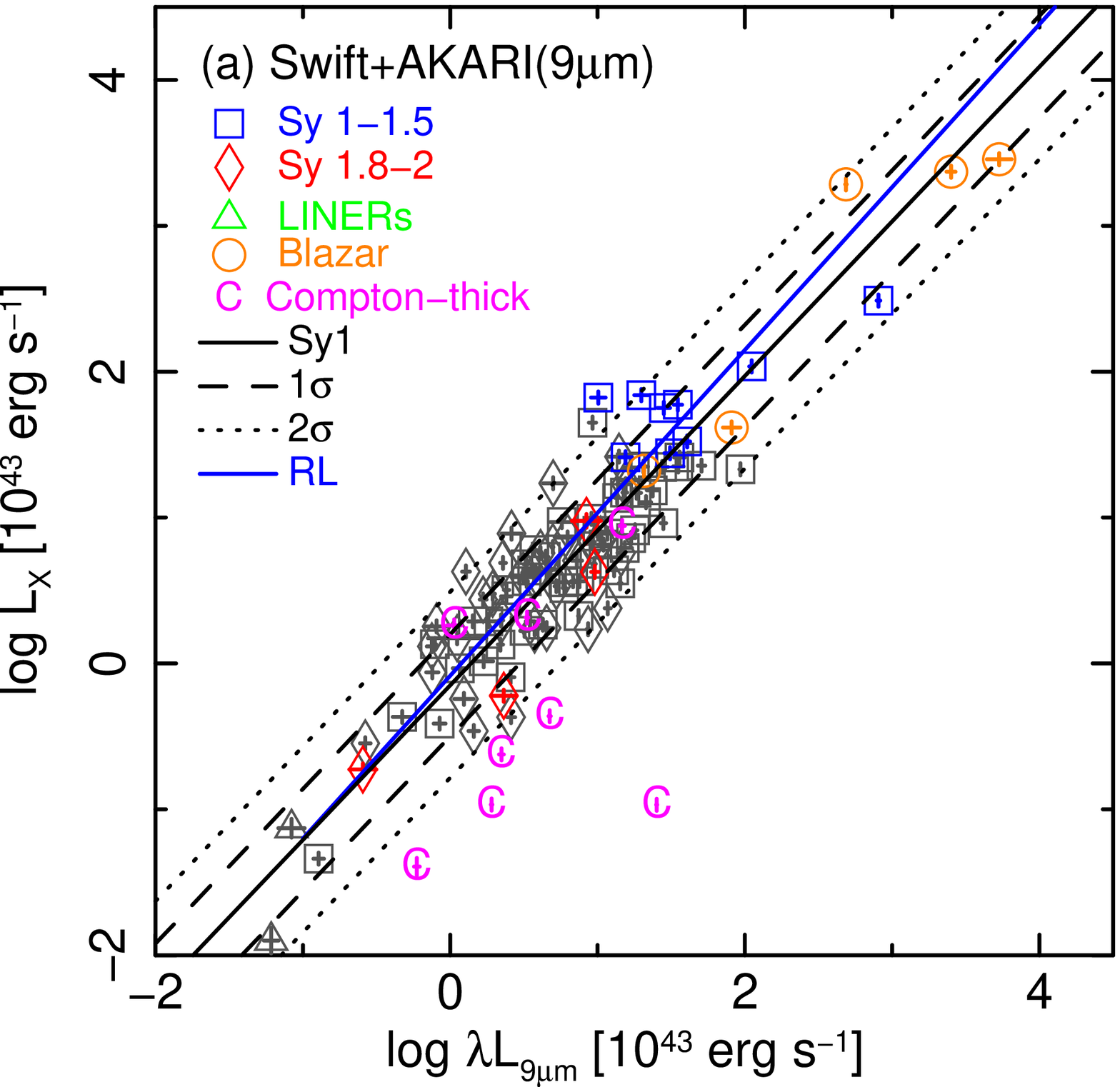}
  \end{center}
 \end{minipage}
 \begin{minipage}{0.33\hsize}
  \begin{center}
   \includegraphics[width=57mm]{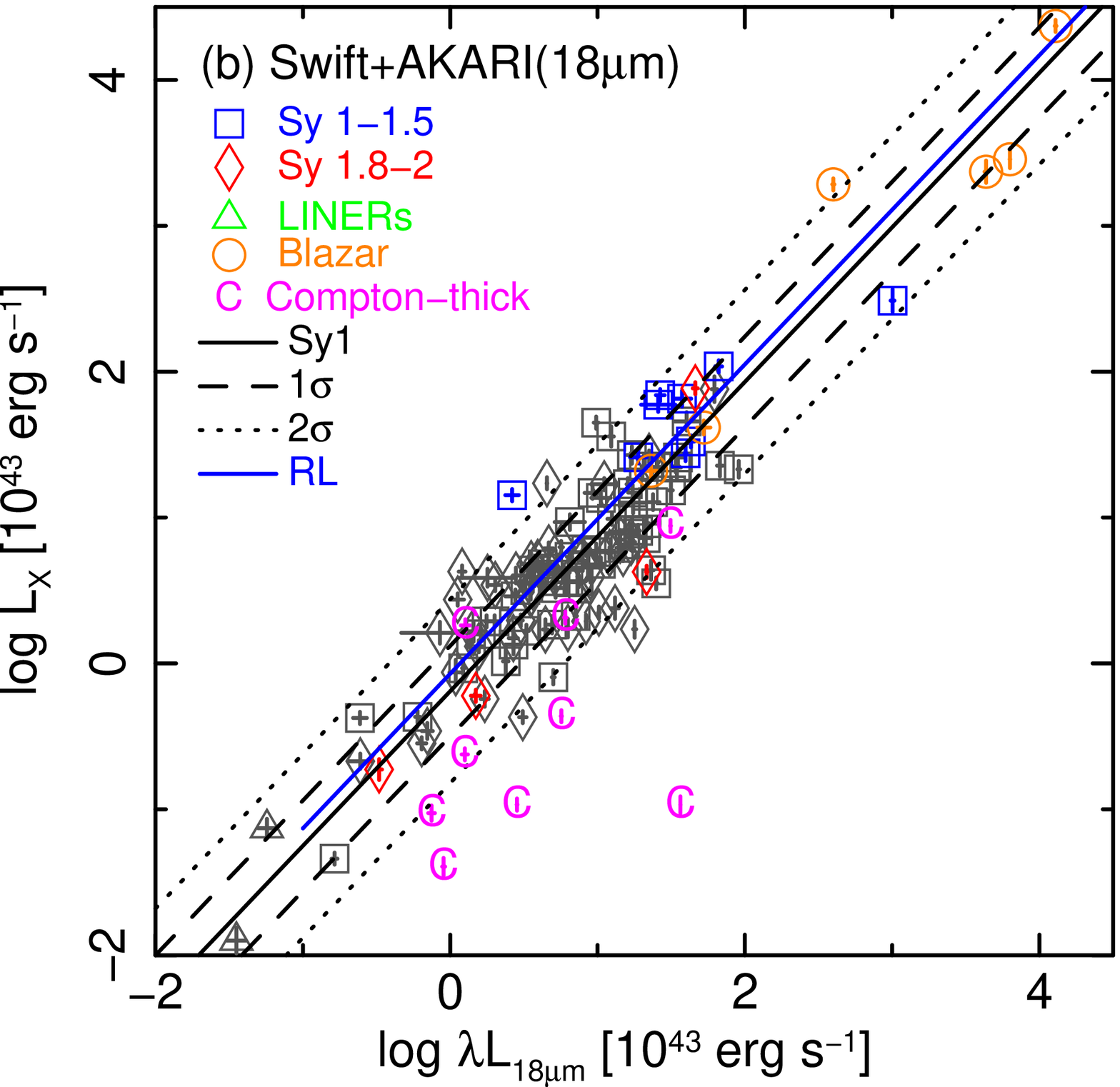}
  \end{center}
 \end{minipage}
 \begin{minipage}{0.33\hsize}
  \begin{center}
   \includegraphics[width=57mm]{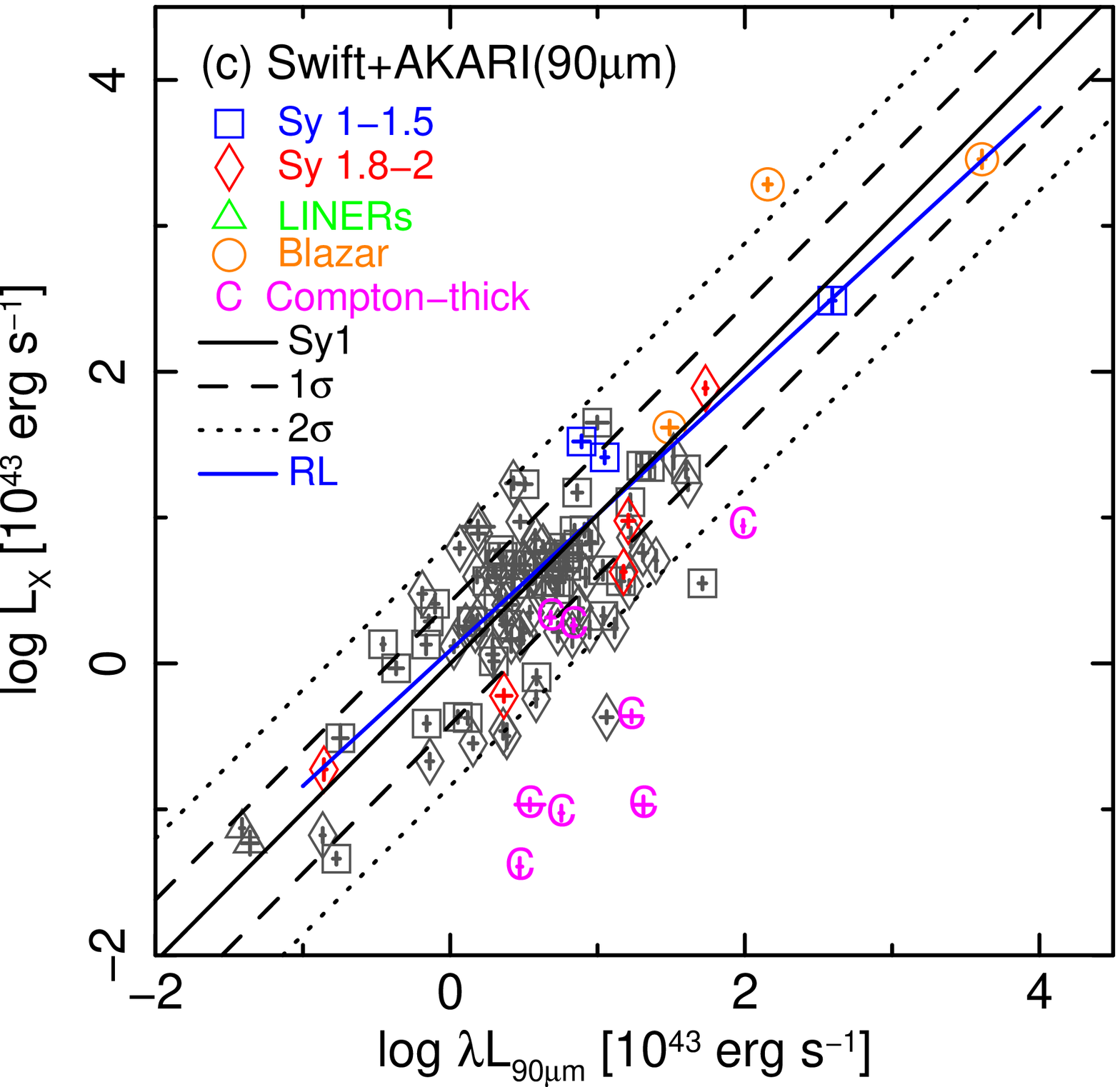}
  \end{center}
 \end{minipage}
\caption{Logarithmic luminosity correlations between the {\it AKARI} and {\it Swift}/BAT.
The ordinates are hard X-ray (14-195~keV) luminosities,
and the abscissas are IR luminosities at (a) 9~$\mu$m, (b) 18~$\mu$m, and (c) 90~$\mu$m.  
Symbols indicates the AGN types (Seyfert~1, Seyfert~2, LINERs, and Blazars). 
Color symbols in the plots are radio-loud AGN and blazars.
Gray symbols are radio-quiet AGN.
Magenta character ``C'' indicates Compton thick sources.
The black and blue solid lines are the correlation fitted to only Seyfert 1s (``Sy1'') and ``RL'', 
by using the liner regression algorithm, respectively.
The dashed and dotted lines show the standard deviation of the correlation; 1$\sigma$ and 2$\sigma$. }
\label{fig:LLplot}
\end{figure}

Based upon a detailed comparison of the complete, 
flux-limited {\it AKARI} and {\it Swift}/BAT all-sky surveys,
we have found a good linear correlation for Seyfert AGN between the logarithms of 
the observed mid-IR and hard X-ray luminosities
over four orders of magnitude (Figure~\ref{fig:LLplot}).
The correlation persists despite the fact that no corrections have been performed for obscuration or reddening, 
nor for any contamination by host galaxy emission. 
The selected surveys thus provide insight into the nature of mid-IR and X-ray populations of sources selected 
from all-sky data in a completely unbiased fashion. 
Statistical tests accounting for redshift effects show that there is definite intrinsic correlation
in the two energies for various classes of AGN.

Hard X-ray emission is produced very near the black hole.
The soft-to-hard X-ray emission of Seyfert AGN is produced by successive Compton 
scatterings of the thermal UV photons emitted by accretion disks in hot, most likely patchy coronae formed above the disks 
\citep[e.g.,][]{1993ApJ...413..507H}. 
The high-energy disk and corona emission also heat dusty tori located at further distances from the centers, 
where the nuclear UV-to-X-ray continuum is re-processed into the thermal IR photons. 
It is widely accepted that the observed radiative output of Seyfert galaxies at mid-IR frequencies is dominated by the torus emission, 
which, at least in the case of unobscured (type 1) sources, may even extend up to the near-IR range 
due to the hot dust located near the dust sublimation radius at the outer edge of the accretion disk \citep[e.g.,][]{2008NewAR..52..274E}.
In this scenario, our result of the good linear correlation for Seyfert AGN between the observed mid-IR and hard X-ray represents well
the view of the unified schemes for Seyfert AGN \citep{1993ARA&A..31..473A}. 

The best-fit correlation lines for Sy1s 
(where we have a direct view of the nuclear regions in both X-rays and IR) are plotted in Figure~\ref{fig:LLplot}
 and are used as the benchmark for comparison again among various source classes in the following sections. 
It is important to note that the correlation when fitting to the sample of ``All'' AGN, is quite similar to the benchmark ``Sy1'' sample, 
which means that these correlations provide an easy route for conversion between IR and hard X-ray powers irrespective of source class. 
If the type of some newly-identified AGN were not known, 
direct use of the best fit parameters for the ``All'' sample could still be used to convert between X-ray and IR powers, 
to within the accuracy provided by the correlation scatter. 

The dispersion in these correlations is $\sim$0.3--0.5 dex (cf. $\sigma_r$ values in Table~\ref{tb:LLfit}). 
The strength of the correlation decreases towards longer wavelengths (lower $\rho$ and $\rho_{\cdot z}$) 
in conjunction with increasing $\sigma_r$ values. 
This may be explained by an increasing host galaxy contribution with wavelength, resulting in a large IR spread. 

It should be noted that although dominated by torus emission in the mid-IR, {\it AKARI}'s beam will encompass a stronger host galaxy contribution as opposed to observations with large ground-based telescopes. This manifests itself as slightly flatter best-fit correlation slopes as compared to the previously-measured correlation using VLT data by \citet[][with the caveat that their sample was much smaller]{2009A&A...502..457G}. 
The slopes we now find for the non-CT radio quiet Seyferts in Table~\ref{tb:LLfit} range over {\it b} (9~$\mu$m) $\approx$ 0.92--0.99, 
as compared to the previously measured value of {\it b} = 1.02 using the same OLS bisector fit statistic. 
The host galaxy contribution is relatively stronger for lower luminosity AGN, 
as a result of which their total observed {\it AKARI} fluxes are systematically biased towards the right (to higher IR powers) in Figure~\ref{fig:LLplot}, 
thus flattening the correlation \citep{2010MNRAS.402.1081V, 2011MNRAS.414.1082M}. 
The effect is small at luminosities of $L_{\rm X}> 10^{43}$~erg s$^{-1}$ and above where most of the BAT AGN lie 
\citep[see, e.g., Eq. 1 of][]{2010MNRAS.402.1081V}, hence the marginal difference in slopes. 
We do not subtract any star formation contribution here because this introduces model dependencies, 
and our aim is to define the correlation properties and source classification diagnostics in {\em observed} space, 
such that they may be useful for future surveys for unclassified objects. Ideally, 
an all-sky infrared survey with 8--10~m telescopes ought to be carried out in the near future; 
this would combine the best angular resolution with complete sample selection. 

In summary, a significant correlation is observed between the hard X-ray and various IR bands for Sy1 and Sy2. 
The correlation is intrinsic to the source fluxes and is a useful empirical tool for converting between the two energies. 


\subsection{Mid-infrared to hard X-ray correlation\\for Compton-thick AGN}\label{discuss:CT-AGN}

The only class of AGN which does not show any correlation between the two bands is that of CT AGN, 
with $\rho_{\cdot z}$ values near zero (Table~\ref{tb:LLfit}). 
Including CT sources within the class of ``Sy2" results in a decrease of the Sy2 correlation as well.
We referred to Table~8.1 in \citet{2004ASSL..308..245C} and other recent results about CT AGN, 
preferring those based upon observations carried out above 10~keV with {\it Beppo}SAX, {\it Suzaku}, and  {\it Swift}/BAT. 
We identified nine CT AGN from our sample. 
Their properties and relevant literature are listed in Table~\ref{tb:compt}.   

\citet{2009A&A...502..457G} showed that 
the mid-IR-to-X-ray correlation for CT AGN is almost indistinguishable from that of typical Seyferts, 
when the X-ray luminosities are corrected for obscuration. 
The difference here is that 
we are using observed hard X-ray fluxes and converting them to luminosities without absorption-correction. 
While absorption does not affect the {\it Swift}/BAT energy band for Compton-thin column densities, 
down scattering does deplete hard X-ray photons when the column becomes CT
\citep[e.g.,][]{2009ApJ...692..608I}. 
Our sample contains a range of obscuring column densities, 
including one AGN with an extreme column of $N_{\rm H}>10^{25}$~cm$^{-2}$ (the CT sources are listed in Table~\ref{tb:compt}). 
This leads to an absence of any correlation for our relatively-small sample of CT AGN.
As we discuss in the next section, the hard X-ray deficit is useful for isolating CT sources.

\subsection{Mid-infrared to hard X-ray correlation\\ for radio-loud AGN}\label{discuss:LLpot}
\subsubsection{Radio-loud AGN}
RL AGN follow a good linear correlation
between the logarithms of mid-IR and hard X-ray luminosities (Figure~\ref{fig:LLplot}).
A correlation is also present in flux--flux space as seen by the high $\rho_{\cdot z}$ (= 0.89)  
for the ``RL'' sample in Table~\ref{tb:LLfit} in all three {\it AKARI} bands. 
The correlation slopes of the RL sample match those of Sy~1s within 1$\sigma$ uncertainties. 
The above similarity suggests 
that the hard X-ray and IR emission processes of RL AGN are similar to Seyferts.
RL AGN probed by {\sl Swift}/BAT lie at higher redshifts than Seyferts ($z_{\rm med}\approx 0.052/0.019$).
But their flux distributions, and in particular, the infrared--to--X-ray flux ratios, 
are very similar to Seyferts ($R_{\rm med}=1.0\pm 0.1/1.2\pm 0.2$), where $R={\rm flux}_{\rm 9 \mu m}/{\rm flux}_{\rm X-ray}$. 

The scatter in the correlation, on the other hand, is larger for the RL sample in all bands. 
This can be seen in Figure~\ref{fig:LLplot} as resulting from a few outliers. In all bands, 
the sources with the largest offset towards higher X-ray luminosities 
(i.e., those lying {\em above} the correlation) are type 1 RL AGN or blazar sources. 
The sources that lie above the 2$\sigma$ uncertainty of the luminosity correlation 
for ``Sy1'' are radio galaxies with particularly powerful jets 3C~111.0, 4C~50.55, 
and radio-loud quasar 3C~273 in 9, 18, and 90~$\mu$m, respectively.

The origin of the broadband emission of radio galaxies is a matter of considerable debate.
With regard to the origin of X-ray emission, 
\citet{2006ApJ...642...96E} demonstrate 
that both accretion and jet-related components may be present in all radio galaxy nuclei, 
with the larger contribution of accretion in more luminous population and vice-versa. 
\citet{2009MNRAS.396.1929H} found a good correlation between mid-IR (15~$\mu$m, {\it Spitzer}) 
and accretion-related X-ray luminosities (in 2--10~keV, {\it Chandra} and {\it XMM-Newton} bands)
in their sample of 135 radio galaxies. 
This suggests that the mid-IR emission is mainly from dusty torus and is not strongly affected by jet beaming, 
especially in the case of luminous sources with intrinsic X-ray powers above $\sim 10^{43}$ erg s$^{-1}$ 
similar to the range of luminosities that we are probing. 
This is also supported by mid-IR spectroscopic analysis of luminous RL AGN \citep{2010ApJ...717..766L}, 
and finally, is also borne out by the similar mid-IR appearance of radio-quiet and radio-loud quasars 
in the composite spectra assembled by \citet{1994ApJS...95....1E}.

\citet{2009MNRAS.396.1929H} did note that 
quasars and BLRG in their sample tended to have higher X-ray luminosities than the low-power radio galaxies of similar IR luminosity, 
suggesting that the X-ray powers of brighter sources can be more severely contaminated by jet emission. 
Our observed outliers appear to show similar X-ray excesses, so 
we examined published detailed X-ray spectroscopic results for these AGN to gain further insight. 
The X-ray spectra of 3C~111.0 obtained with {\it Suzaku} 
indeed showed jet dominant component in the hard X-ray band, as found by \citet{2011arXiv1108.2609B}.
On the other hand, 
in the broadband X-ray spectra of 4C~50.55 determined by {\it Suzaku} and {\it Swift}/BAT data,
there is little jet contribution to hot coronal Comptonization \citep{2010ApJ...721.1340T}. 
3C~273 will be discussed in the next section. 
It is interesting to note that both 3C~111.0 and 4C~50.55 belong to the class of BLRG, 
which was the only class for which \citet{1999ApJ...526...60S} 
identified some systematic differences in X-ray spectral slope (albeit weak) with respect to radio quiet AGN. 
On the other hand, 
other BLRGs such as 3C~120 \citep{2007PASJ...59..279K} are not largely offset from our correlation. 

In summary, 
we lack conclusive evidence of any strong and systematic differences for our RL samples as a whole, though some jet contribution (and related excess variability) may explain slight offsets in the IR--X-ray correlation plane for some objects. 
In any case, RL sources join on to the luminous end of the sample of Seyferts and extend the observed correlation in that regime. 

\subsubsection{Blazars}\label{discuss:LLpot-blazar}
The broadband spectra of blazars are thought to be dominated by non-thermal emission from their relativistic jets 
\citep{1990A&ARv...2..125B, 1998MNRAS.299..433F}. 
The IR regime of FSRQs is mainly synchrotron radiation, 
and hard X-rays are from the inverse-Compton scattering of soft target photons produced 
either within the jet (synchrotron self-Compton process) or external to the jet (e.g., within the broad emission line region or dusty tori).
In the case of low-power BL Lacs, 
X-rays are typically dominated by the high-energy tail of the synchrotron continuum. 
Such ``high frequency-peaked'' BL Lacs are however absent in our BAT/{\it AKARI} sample.
Other components including accretion disk and host galaxy emission are present 
in most sources at a weaker level \citep[e.g.,][]{2010ApJ...716...30A}. 
Whether or not a torus contributes to the IR emission is debated, 
but it is generally found to be present more often in luminous FSRQ sources than in BL Lac
\citep[e.g.,][]{2011ApJ...732..116M, 2011arXiv1112.5162P}. 

Our blazar sample contains four FSRQs and two BL Lac.
Although the sample size is small, 
we find high values for partial correlation coefficients between the IR and hard X-ray fluxes for blazars 
($\rho_{\cdot z} \approx 0.8-0.9$ in 9 and 18~$\mu$m; see Table~\ref{tb:LLfit}). 
This is expected if the various synchrotron and inverse Compton components are intrinsically correlated. 
On the other hand, the apparent observation from Figure~\ref{fig:LLplot} that blazars extend the Sy1 correlation is likely to be a coincidence. 
This is because the correlation implies a near 1:1 absolute normalization 
(i.e., mid-IR luminosities are very similar to the hard X-ray powers), 
which may be explained according to the \lq blazar sequence\rq\ of \citet{1998MNRAS.299..433F}. 
Figure~12 of their work shows that 
the {\it AKARI} (9~$\mu$m) and {\it Swift}/BAT (14--195~keV) bandpasses sample similar power levels at all luminosities, 
to within a factor of a few at most. 
Furthermore, Malmquist bias certainly contributes to the luminosity-to-luminosity correlation 
in flux-limited samples for objects spanning a limited flux range but a large redshift range \citep[e.g.,][]{2011arXiv1101.0837A}, 
which is true in the case of BAT-selected blazars. 
A larger flux range needs to be sampled before detailed conclusions on the origin and validity of the correlation for blazars may be drawn. 

Amongst blazars, 
the FSRQ 3C~273 shows the strongest mismatch between the observed hard X-ray and IR powers 
(lying on or well above the 2$\sigma$ Sy correlation lines in Figure~\ref{fig:LLplot}). 
This object is known to show strong multi-component variability \citep[e.g.,][]{1987A&A...176..197C}, 
which may account for its position in the IR-to X-ray plane.

\vspace{10pt}
\noindent
In summary, 
results from the above two sections imply that our IR-to-X-ray correlation may be used to predict and convert
between the mid-IR and hard X-ray powers of RL AGN, in addition to Seyferts. 
The hard X-ray/IR-brightest blazars included in our flux-limited sample happen to lie along this correlation as well, 
which is likely to be a coincidence and not an indication for the presence of dust. 
Thus, the IR vs. X-ray correlation is a useful empirical tool irrespective of the underlying emission physics. 
This is confirmed by the fact that when we gather 
all Compton-thin AGN together, irrespective of radio classification, 
large positive values of $\rho_{\cdot z}$ are found 
(see ``All, ex CT'' subsample in Table~\ref{tb:LLfit}).

%

\subsection{Photometric diagnostics for source classification}

Photometric diagnostics for identifying various source classes 
(e.g., by using observed power or flux ratios between various bands) can be simple and powerful observational tools. 
Unbiased all-sky surveys such as ours are ideal for testing various classification schemes,
which may be applied to large blind surveys, and we have carried out such a search. 
The IR-to-X-ray correlations for Sy1, Sy2, and LINERs closely match each other
\citep{2009A&A...502..457G, 2011arXiv1109.4873A}. 
Here, we investigate to what extent Seyferts, RL AGN, and CT sources may be separated 
using the mid-IR and hard X-ray correlations.

\subsubsection{Radio-loud AGN}\label{discuss:abs-color}
\begin{figure}[!ht]
\begin{center}
   \includegraphics[width=100mm]{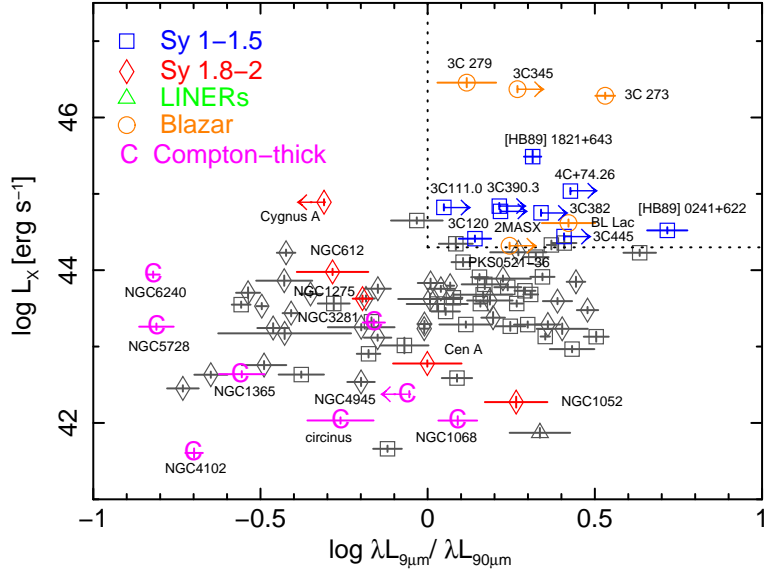}
\end{center}
\caption{Absolute magnitude (log $L_{\rm X}$) vs. 
Color (log $\lambda L_{9\mu {\rm m}}/\lambda L_{90\mu {\rm m}}$) diagram
for 89 {\it Swift}/BAT and {\it AKARI} detected sources.
Color symbols in the plots are radio-loud AGN and blazars.
Gray symbols are radio-quiet AGN.
Magenta character ``C'' indicates Compton thick sources.
The radio loud sources are located in the upper-right region as indicated by the dotted box.
NGC3079, one of the Compton-thick sources, is out of the figure range, (log $L_{\rm X}$, log $\lambda L_{9\mu {\rm m}}/\lambda L_{90\mu {\rm m}}$) = (41.93, -1.69).}
\label{fig:hardness-ratio-radio}
\end{figure}

As already mentioned, RL AGN (and blazars, with the aforementioned caveats) follow the same correlation as Seyferts  in Figure~\ref{fig:LLplot}. 
But these sources lie at the high luminosity end on both axes. 
RL AGN  can also be separated 
by using a luminosity vs. flux ratio ($L_{\rm X}$ vs. $\lambda L_{\lambda(9\mu {\rm m})}$/$\lambda L_{\lambda (90\mu {\rm m})}$) plot,
equivalent to an absolute magnitude vs. color diagram. 
The result is shown in Figure~\ref{fig:hardness-ratio-radio}. 
Non-detected RL AGN were included by using IRC 9~$\mu$m and FIS 90~$\mu$m detection limits of 0.12 and 0.55~Jy, respectively.  

Most Seyferts show a large spread in the mid-IR-to-far-IR ratio with an average value of 
$\lambda L_{\lambda(9\mu {\rm m})}$/$\lambda L_{\lambda (90\mu {\rm m})}$ = 1.03. 
Type 1 RL AGN and blazars, 
on the other hand, occupy the region of high hard X-ray emission and high mid-IR-to-far-IR ratio as compared to other AGN types. 
Using approximate boundaries of $\log L_{\rm X} > 44.3$ and 
$\lambda L_{\lambda(9\mu {\rm m})}$/$\lambda L_{\lambda (90\mu {\rm m})}>1$ for the box shown in the figure, 
we find that {\em all} type 1 RL AGN and blazars lie in this region, i.e., selection is 100\% {\em complete}. 
On the other hand, the box also contains a few (2/16) sources that are neither type 1 RL AGN nor blazars, 
i.e., selection is 88\% {\em reliable}. 
If we classify [HB89]0241+622 and [HB89]1821+643 as radio-quiet (see discussion in section~\ref{sec:sourceid}),
the reliability drops to 75\%.

On the other hand, type 2 RL AGN are not clearly separated in this figure. 
In particular, 
they show a lower ratio of the mid- to far-IR powers on the ordinates-to-the median values 
for $\lambda L_{\lambda(9\mu {\rm m})}$/$\lambda L_{\lambda (90\mu {\rm m})}$ for type 1 RL and type 2 RL are 0.31 and --0.11, respectively. 
This trend is also seen in the case of the radio-quiet sources (RQ), 
with the corresponding ratios being 0.16 and --0.16 for type 1 RQ and type 2 RQ, respectively. 
This trend cannot arise from optically-thick tori in Sy2 being preferentially \lq colder\rq\ in the infrared. 
If this were so, then Sy2 would also be offset towards lower 9~$\mu$m--to--X-ray flux ratios in the IR-to-X-ray correlations. 
Instead, this trend probably arises 
because Sy2 have lower intrinsic accretion (X-ray) luminosities as compared to Sy1 on average \citep{2009ApJ...690.1322W}, 
as well as a relatively stronger 90~$\mu$m contribution from dust in the host galaxy \citep{1998ApJS..117...25M}. 
These two facts push down the 9~$\mu$m fluxes for Sy2 and boost their 90~$\mu$m fluxes, resulting in the observed trend.

With regard to the fact that blazars occupy the top right of Figure~\ref{fig:hardness-ratio-radio}, 
an x-axis mid--to--far-IR luminosity ratio greater than 1 means 
that their jet synchrotron peaks must lie in the mid-IR or higher frequencies and not the far-IR or lower frequencies.
This is interestingly consistent with the recent study of bright blazars detected at GeV photon energies by {\it Fermi}/LAT \citep{2010ApJ...716...30A}, 
where no source with the synchrotron continuum peaked at far-IR or lower frequencies was found.
Further investigation is beyond the scope of this paper and is left to future work. 

Finally, it is seen from Figure~\ref{fig:hardness-ratio-radio} that 
there are no luminous X-ray sources with a mid-IR-to-far-IR ratio of less than 1 occupying the top left of the diagram. 
Type 2 quasars with powerful X-ray emission and bright host galaxies probably occupy this region, 
but these are known to be elusive and absent in the local universe 
\citep[e.g.,][]{2003AJ....126.2125Z, 2004MNRAS.348..529G, 2006MNRAS.369.1566G}. 
We also note that we would not have selected sources 
that are below the {\it AKARI} detection limit in both 9 and 90~$\mu$m. 
Since our sample contains very few blazars, 
a selection bias where we miss such sources cannot be ruled out. 

In summary, Figure~\ref{fig:hardness-ratio-radio} is helpful for isolating type 1 RL AGN and blazars. 
Using color and luminosity information we can isolate type 1 RL AGN with high reliability, 
but this requires knowledge of source redshift.

\subsubsection{Compton-thick AGN}
Outliers in the correlation plots of Figure~\ref{fig:LLplot}, which lie {\em below} the correlation for ``Sy1'', are mostly CT AGN. 
Their average $L_{\rm X}$/$\lambda L_{\lambda(9\mu {\rm m})}$ ratio is 0.11, as compared to 0.71 for Sy1. 
When generating such correlation plots, 
it is important to use the observed X-ray flux for identifying CT AGN
because of the hard X-ray deficit due to Compton down-scattering (Sec.~\ref{discuss:CT-AGN}).

The average values of the X-ray-to-IR ratios for other {\it AKARI} bands are 
$L_{\rm X}$/$\lambda L_{\lambda(18\mu {\rm m})} = 0.08$
and $L_{\rm X}$/$\lambda L_{\lambda(90\mu {\rm m})} = 0.03$. 
Copious star formation often accompanies obscured AGN activity and several of the CT AGN 
lie in composite systems also classified as starburst galaxies 
(e.g., NGC~6240, Circinus Galaxy, NGC~4945. See Table~\ref{tb:compt}). 
The 90~$\mu$m to X-ray ratio for CT AGN is large because of an excess far-IR contribution 
due to bright star formation in the host galaxy (see also Figure~\ref{fig:LLplot}). 

\subsubsection{Distinguishing between Compton-thick AGN and Starburst galaxies}\label{discuss:ratio}
The difficulty of distinguishing heavily obscured AGN in composite systems from pure starburst galaxies has been an important problem for many decades
\citep[e.g.,][]{1998ApJ...498..579G, 1998ApJ...505..174G, 2006ApJ...649...79S}. 
X-ray information only below 10~keV has been used in most surveys so far. 
But the observed X-ray fluxes for both these classes of sources in this band are low as compared to their observed IR fluxes. 
Thus, such sources are often indistinguishable photometrically. 
High excitation mid-IR forbidden lines and PAHs can separate the two 
\citep[e.g.,][]{2000A&A...359..887L, 2002A&A...393..821S, 2006ApJ...646..161D, 2009MNRAS.398.1165G, 2010ApJ...716.1151W},
if spectroscopic data are available. 

We now have new information available from the hard X-ray band above 10~keV.
We investigate whether the information is useful for source classification. 
We checked the hard X-ray emission of starburst galaxies. 
There is no starburst galaxy detected in the 22-month BAT catalog, 
so we have extended our search to the 54-month BAT catalog \citep{2010A&A...524A..64C} for this purpose. 
The detection limits of the 54-month catalog are 1.0$\times$10$^{-11}$ ($|b|<10^\circ$)
and  9.2$\times$10$^{-12}$ erg cm$^{-2}$ s$^{-1}$ ($|b|>10^\circ$), respectively. 
M82 is the only ``pure'' starburst galaxy detected in this catalog.

Though only one starburst galaxy is detected, hard X-ray predictions on other well-known sources turn out to be interesting for our purpose of comparison with AGN. We picked up two famous ``pure'' starburst galaxies (NGC~253, Arp~220), 
and other four ``pure'' starburst galaxies (NGC~2146, NGC~3256, NGC~3310, and NGC~7714) 
listed in \citet{2006ApJ...653.1129B}, which are all detected in X-rays below 10~keV. 
We computed hard X-ray flux estimates for these 6 objects by 
using their observed 2--10~keV fluxes under the two different assumptions of non-thermal and thermal radiation models 
(The results are listed in Table~\ref{tb:sbg}).  
These correspond to the emission expected from point sources like X-ray binaries 
and diffuse thermal gas, respectively \citep{2002A&A...382..843P}.
The non-thermal model is based upon the observation of star-forming galaxies in the Hubble Deep Field with an average 2--10~keV X-ray spectral slope (photon index $\Gamma = 2.1$; \citealt{2003A&A...399...39R}). 
The thermal model (7~keV {\tt APEC}) is based on the {\it XMM-Newton} observation of the central region of M82 \citep{2008MNRAS.386.1464R}.
Similar starburst-related hot gas bubbles have also been found in Arp 220 \citep{2005MNRAS.357..565I} and NGC~253 \citep{2001A&A...365L.174P}. 

Because M82 and NGC~253 are very bright in the IR ($\ge100$~Jy),
the {\it AKARI} detectors may cause saturation and the flux could have larger uncertainty \citep{2010yCat.2298....0Y}. 
The FIS 90~$\mu$m fluxes of these two sources are not reliable ({\it FQUAL} = 1), 
and there is no IRC 9~$\mu$m flux available for M82 in the all-sky survey. 
Therefore we applied the results of the {\it AKARI} pointed observation data 
for M82 \citep{2010A&A...514A..14K} and NGC~253 \citep{2009ApJ...698L.125K}. 
In these observations, the IRC and FIS were operated in special observation mode,
which is used to observe bright sky regions to avoid detectors saturation.

\begin{figure}[!ht]
\begin{center}
      \includegraphics[width=100mm]{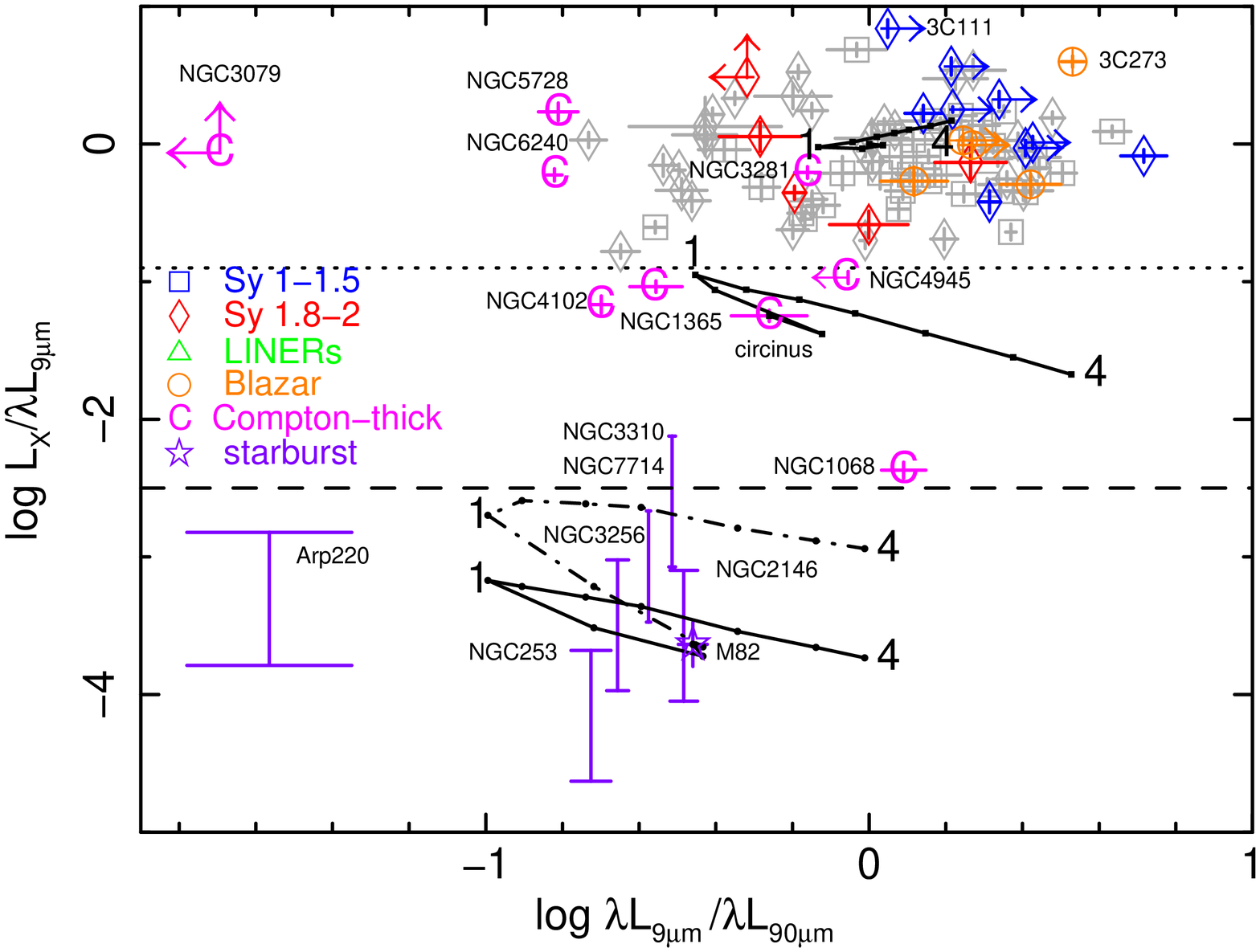}
   \end{center}
\caption{Color-color plot of 
log $L_{\rm X}$/$\lambda L_{9\mu {\rm m}}$ vs. 
log $\lambda L_{9\mu {\rm m}}$/$\lambda L_{90\mu {\rm m}}$ for 89 AGN
including 9 Compton-thick (CT) sources and 7 starburst galaxies detected by {\it AKARI}.
Color symbols in the plots are radio-loud AGN and blazars.
Gray symbols are radio-quiet AGN.
Magenta character ``C'' indicates Compton thick sources.
Open purple stars show starburst galaxies in Table~\ref{tb:sbg},
and indicate the estimated hard X-ray (14--195~keV) flux ranges
by using the non-thermal (higher bar) and thermal model (lower bar).
The dotted line at log $L_{\rm X}/\lambda L_{9\mu {\rm m}}\sim$ --0.9 shows 
an approximate boundary between normal AGN and CT sources.
The dashed line at log $L_{\rm X}/\lambda L_{9\mu\mathrm{m}} \sim -2.5$
shows the boundary below which only starburst galaxies are present.
The solid lines show how the sources move on the color-color plots when the redshift is changed from 0.0 to 4.0.
The lines are calculated for three types of galaxies, 
a Seyfert galaxy located at (0, 0),  
Circinus galaxy as a CT AGN,
and M82 as a starburst galaxy.
The solid and dot-dashed lines for M82 correspond to the different model (non-thermal and thermal, respectively).}
\label{fig:l90vsl9-lxvsl9-sbg-estimate}
\end{figure}

Figure~\ref{fig:l90vsl9-lxvsl9-sbg-estimate} shows the results in the form of a 
$\log L_{\rm X}/\lambda L_{\lambda(9\mu {\rm m})}$ vs. 
$\log \lambda L_{\lambda(9\mu {\rm m})}/\lambda L_{\lambda(90\mu {\rm m})}$ diagram (equivalent to a color--color plot). 
We also include an IR color lower limit for the CT AGN, NGC~4945, resulting from it being saturated in the FIS ($\sim$ 100~Jy). Other CT AGN non-detected at 9~$\mu$m are also included as limits. 
Considering the AGN only, 
an approximate boundary of $\log L_{\rm X}/\lambda L_{\lambda(9\mu {\rm m})} <$ --0.9 successfully isolates CT AGN, 
i.e., it has very high (100\%) reliability. 
But only 5/8 of CT sources are below this boundary, 
which means that it is 56\% complete. 
This boundary can successfully identify CT sources, if a source is known to contain an AGN a priori.

But such a priori information is not available for most surveys, 
and we see from the figure that starburst galaxies also occupy this region. 
Starburst galaxies (including the single detection of M82 and virtually all prediction ranges) are all concentrated below the line of 
$\log L_{\rm X}/\lambda L_{\lambda(9\mu\rm{m})}\sim-2.5$. 
For each source, 
a vertical line connects the expected range of values based upon the non-thermal model for the X-ray emission (higher horizontal bar) 
and the thermal model (lower horizontal bar). 
This shows that the hard X-ray emission from starburst galaxies is expected 
to be less than $\sim$ 10$^{-3}$ of the mid-IR emission from AGN. 

The abscissa, $\log \lambda L_{\lambda(9\mu {\rm m})}/\lambda L_{\lambda(90\mu {\rm m})}$, 
shows the ratio of hot-to-cool dust.
Starburst galaxies are largely separated on the abscissa as well because of cooler dust temperature compared to AGN. 
We also tested the ratio of 15 other ``pure'' starburst galaxies listed in \citet{2006ApJ...653.1129B},
and obtained an upper limit of $\log \lambda L_{\lambda(9\mu {\rm m})}/\lambda L_{\lambda(90\mu {\rm m})}\sim$ --0.4. 
The result shows that
the far-IR emission from the cool dust is more dominant than the mid-IR emission from hot dust in the local starburst galaxies.


On the other hand, RL AGN are not separated out in this figure. 
3C~111 and 3C~273 lie towards the top right region in the plot (again 3C~111 has the largest  $\log L_{\rm X}$/$L_{\rm IR}$ ratio), 
so it might be possible to define some region here,
but in general RL AGN and blazars do not occupy any specific region.

In summary, the hard X-ray and IR color--color plot successfully differentiates  the composite CT AGN/starburst galaxies, and also distinguishes these 
from the remaining sources (Sy1, Sy2, LINERs, and RL AGN). 

\subsubsection{{\it K}-corrections in the color--color plane}
In order to be useful for general source classification, 
the regions determined in the previous section must be shown to accommodate the effect of SED {\it K}-corrections at higher redshifts. 
We checked this 
by applying {\it K}-correction for template SEDs of Seyferts, CT AGN, and starburst galaxies separately. 
For Seyferts, we use the high-luminosity starburst model spectrum ($L = 10^{13} L_\odot$) by \citet{2003MNRAS.338..555L}.
For CT AGN and starburst galaxies, we use the Seyfert 2 and M82 model spectra, respectively, by \citet{2007ApJ...663...81P}. 
We show the color tracks on the plane of Figure~\ref{fig:l90vsl9-lxvsl9-sbg-estimate} when the redshift is varied from $z=0.0$ to 4.0.
We assumed a color--color position of (0, 0) for the local origin of the {\it K}-correction locus for Seyferts.
Similarly, the loci of CT AGN and starburst galaxies start at the positions of Circinus Galaxy and M82, respectively.
The plotted loci show that our diagnostic regions are independent of source redshift.

With current and upcoming hard X-ray missions of continually improving sensitivity
(e.g., {\it INTEGRAL}, {\it Swift}, {\it NuSTAR}, {\it eROSITA}, and {\it ASTRO-H}), 
and similar increases in sensitivity expected in the IR (e.g., {\it WISE} and {\it SPICA}),
our work investigates the degree to which blind 
IR/X-ray surveys can be used for classification and separation of AGN and starburst galaxies, 
using directly observed photometric properties without the requirement of detailed spectroscopy or source modeling. 
For instance, changing from the {\it Swift}/BAT 14--195~keV and {\it AKARI} 9~$\mu$m bands 
to the {\it NuSTAR} 6--79~keV and {\it WISE} 12~$\mu$m bands, respectively, 
the maximal shift in the ordinates of the color--color plane is only $\sim$20\%  towards lower $L_{\rm X}$, 
for a Seyfert SED at $z \sim 4$.


\section{Conclusions}
We have combined two complete, flux-limited all-sky surveys, 
{\it AKARI} and {\it Swift}/BAT,
in order to study the connection between the IR and hard X-ray ($> 10$~keV) bands for AGN.
We found a good linear correlation between the logarithms of the observed mid-IR and hard X-ray luminosities 
over four orders of magnitude.
Under the AGN  orientation-based unification scheme, this result may be viewed as supporting clumpy torus models,
although the broad-band correlation alone cannot prove the underlying emission mechanism 
for the IR and hard X-ray bands.  
We also found that all the RL AGN follow a good linear correlation,
suggesting that their dominant hard X-ray and IR emission processes are similar to those of radio-quiet sources.
Coincidentally, blazars also possess a near 1:1 intrinsic correlation between the two bands, 
which is due to the fact that the synchrotron and inverse Compton powers match each other well along the blazar sequence.
We present quantitative correlation statistics and show 
that artificial redshift effects do not dominate the correlation for Seyferts nor for RL AGN. 
The correlation fits are a useful empirical tool for observers. 
Most CT AGN show a large deficit in their observed X-ray powers compared to unobscured AGN. 
This is expected because their observed fluxes are diminished by high gas column densities, even in the hard X-ray band. 
  
We investigated hard X-ray and IR photometric diagnostics for source classification.
We could isolate type 1 RL AGN and blazars by using absolute magnitude vs. color diagram such as 
$\log L_{\rm X}$ vs. $\lambda L_{\lambda(9\mu {\rm m})}$/$\lambda L_{\lambda(90\mu {\rm m})}$. 
Using approximate boundaries of $\log L_{\rm X}> 44.3$ and 
$\lambda L_{\lambda(9\mu {\rm m})}$/$\lambda L_{\lambda (90\mu {\rm m})}>1$, 
we found that {\em all} type 1 RL AGN and blazars lie in this region with a reliability of 88\%. 
The optical Seyferts classification is related to the 9 vs. 90~$\mu$m infrared flux ratio with type 2 sources showing a lower ratio than type 1. 
We discussed biases resulting from our flux limited selection and 
the absence of rare populations such as type 2 quasars in the present, local sample.

We also found that the color--color plot of 
log $L_{\rm X}$/$\lambda L_{\lambda(9\mu {\rm m})}$ vs. 
log $\lambda L_{\lambda(9\mu {\rm m})}$/$\lambda L_{\lambda(90\mu {\rm m})}$
is able to distinguish starburst galaxies and composite CT sources from other normal AGN. We defined several boundaries, which can be used for reliable and complete source classification. 
CT AGN are successfully isolated by the approximate boundary of $\log L_{\rm X}$/$\lambda L_{\lambda(9\mu {\rm m})} <$ --0.9 
with very high (100\%) reliability amongst AGN, but completeness of 56\%.
Starburst galaxies are isolated in a region with approximate boundary of $\log L_{\rm X}/L_{\lambda (9\mu {\rm m})}<-2.5$ and 
log $\lambda L_{\lambda(9\mu {\rm m})}$/$\lambda L_{\lambda(90\mu {\rm m})}<-0.4$ and very high (100\%) reliability.\\ 

{\small 
We acknowledge the work of the {\it AKARI}/PSC and the {\it Swift}/BAT 22-month Source Catalog teams.  
We also acknowledge  the referee for constructive comments and suggestions.
This work was partly supported by the Grant-in-aid for Scientific Research
20740109, 21244017 (YT), and 23540265 (YU) from the Ministry of Education, Culture, Sports, Science,
and Technology of Japan.}


\begin{table}[!ht]
\begin{center}
\caption{AKARI infrared propaties of the AGN in the 22-month {\it Swift}/BAT catalog}\label{tb:CASresult-detail}
\begin{tabular}{lllrrrccl}\hline\hline
Serial & BAT & Object name &Flux$_{09}$ & Flux$_{18}$ & Flux$_{90}$ & log $L_{\rm X}$ & Redshift & Type\\
NO. & No. & & (mJy) & (mJy) & (mJy) & (erg~s$^{-1}$) &  & \\
(1) & (2) & (3) & (4) & (5) & (6) & (7) & (8) & (9)\\\hline
1 & 1 & Mrk 335 & 128 $\pm$ 3 & 223 $\pm$ 46 & \nodata  & 43.52 $\pm$ 0.07 & 0.0258 & Sy1.2 \\   
2 & 2 & Mrk 1501 & \nodata  & 134 $\pm$ 21 & \nodata  & 44.82 $\pm$ 0.05 & 0.0893 & Sy1.2/RIM \\   
3 & 3 & 2MASX J00253292+6821442 & 75 $\pm$ 8 & 216 $\pm$ 22 & \nodata  & 42.94 $\pm$ 0.09 & 0.0120 & Sy2 \\   
4 & 8 & NGC 235A & \nodata  & 295 $\pm$ 32 & \nodata  & 43.61 $\pm$ 0.05 & 0.0222 & Sy1$^{\rm a}$ \\  
5 & 9 & Mrk 348 & \nodata  & 593 $\pm$ 43 & 736 $\pm$ 60 & 43.79 $\pm$ 0.04 & 0.0150 & Sy2 \\   
6 & 21 & NGC 526A & 141 $\pm$ 14 & 292 $\pm$ 26 & \nodata  & 43.64 $\pm$ 0.04 & 0.0191 & Sy1.5 \\   
7 & 22 & Fairall 9 & 229 $\pm$ 20 & 440 $\pm$ 24 & \nodata  & 44.35 $\pm$ 0.04 & 0.0470 & Sy1 \\   
8 & 23 & NGC 513 & \nodata  & \nodata  & 2533 $\pm$ 107 & 43.20 $\pm$ 0.08 & 0.0195 & Sy2 \\   
9 & 24 & NGC 612 & 135 $\pm$ 34 & \nodata  & 2604 $\pm$ 261 & 43.98 $\pm$ 0.04 & 0.0298 & Sy2$^{\rm a}$/RL \\
10 & 25 & ESO 297-018 & \nodata  & \nodata  & 672 $\pm$ 67 & 43.97 $\pm$ 0.04 & 0.0252 & Sy2 \\   
11 & 27 & MCG -01-05-047 & \nodata  & 139 $\pm$ 7 & 2597 $\pm$ 225 & 43.22 $\pm$ 0.07 & 0.0172 & Sy2 \\   
12 & 28 & NGC 788 & \nodata  & 312 $\pm$ 25 & \nodata  & 43.54 $\pm$ 0.04 & 0.0136 & Sy2 \\   
13 & 34 & NGC 931 & 349 $\pm$ 12 & 763 $\pm$ 48 & 2430 $\pm$ 64 & 43.56 $\pm$ 0.04 & 0.0167 & Sy1.5 \\   
14 & 35 & IC 1816 & 55 $\pm$ 32 & 265 $\pm$ 20 & 1473 $\pm$ 54 & 43.17 $\pm$ 0.07 & 0.0170 & Sy1.8 \\   
15 & 36 & NGC 973 & \nodata  & 93 $\pm$ 77 & 1704 $\pm$ 69 & 43.21 $\pm$ 0.07 & 0.0162 & Sy2 \\   
16 & 37 & NGC 985 & 165 $\pm$ 14 & 368 $\pm$ 38 & 1291 $\pm$ 16 & 44.11 $\pm$ 0.05 & 0.0430 & Sy1 \\   
17 & 41 & NGC 1052 & 146 $\pm$ 34 & 377 $\pm$ 17 & 793 $\pm$ 36 & 42.27 $\pm$ 0.07 & 0.0050 & Sy2/RL \\   
18 & 42 & NGC 1068 & 24923 $\pm$ 188 & 72285 $\pm$ 27 & 202591 $\pm$ 28886 & 42.03 $\pm$ 0.06 & 0.0038 & Sy2/CT \\   
19 & 43 & [HB89] 0241+622 & 300 $\pm$ 14 & 635 $\pm$ 34 & 576 $\pm$ 80 & 44.52 $\pm$ 0.04 & 0.0440 & Sy1/RIM \\   
20 & 44 & 2MASX J02485937+2630391 & \nodata  & \nodata  & 1398 $\pm$ 25 & 44.42 $\pm$ 0.08 & 0.0579 & Sy2 \\   
21 & 47 & NGC 1142 & 265 $\pm$ 16 & 380 $\pm$ 71 & 7036 $\pm$ 178 & 44.23 $\pm$ 0.04 & 0.0289 & Sy2 \\   
22 & 50 & NGC 1194 & 169 $\pm$ 7 & 415 $\pm$ 56 & 530 $\pm$ 42 & 43.13 $\pm$ 0.07 & 0.0136 & Sy1 \\   
23 & 54 & NGC 1275 (3C~084) & 442 $\pm$ 26 & 1989 $\pm$ 21 & 6928 $\pm$ 245 & 43.63 $\pm$ 0.04 & 0.0176 & Sy2/RL \\   
24 & 55 & B2~0321+33 NED02 & \nodata  & 95 $\pm$ 4 & \nodata  & 44.56 $\pm$ 0.07 & 0.0610 & Sy1 \\   
25 & 57 & PKS 0326-288 & \nodata  & 152 $\pm$ 7 & \nodata  & 44.88 $\pm$ 0.09 & 0.1080 & Sy1.9 \\   
26 & 60 & NGC 1365 & 2234 $\pm$ 38 & 5364 $\pm$ 43 & 80384 $\pm$ 13546  & 42.64 $\pm$ 0.04 & 0.0055 & Sy1.8/CT \\   
27 & 64 & ESO 548-G081 & \nodata  & \nodata  & 968 $\pm$ 61 & 43.28 $\pm$ 0.04 & 0.0145 & Sy1 \\   
28 & 68 & 2MASX J03534246+3714077 & \nodata  & \nodata  & 827 $\pm$ 78 & 43.06 $\pm$ 0.10 & 0.0183 & Sy2 \\   
29 & 70 & ESO~549- G049 & 156 $\pm$ 15 & 302 $\pm$ 43 & 2976 $\pm$ 170 & 43.56 $\pm$ 0.09 & 0.0263 & Sy1$^{\rm a}$ \\ 
30 & 72 & IRAS 04124-0803 & 168 $\pm$ 12 & 423 $\pm$ 39 & \nodata  & 43.86 $\pm$ 0.08 & 0.0379 & Sy1 \\   
31 & 73 & 3C 111.0 & 61 $\pm$ 7 & \nodata  & \nodata  & 44.82 $\pm$ 0.04 & 0.0485 & Sy1/RL \\   
32 & 75 & 1H 0419-577 & \nodata  & 106 $\pm$ 21 & \nodata  & 44.66 $\pm$ 0.06 & 0.1040 & Sy1 \\   
33 & 77 & 3C 120 & 203 $\pm$ 20 & 497 $\pm$ 68 & 1468 $\pm$ 85 & 44.41 $\pm$ 0.04 & 0.0330 & Sy1/RL \\   
34 & 78 & MCG -02-12-050 & \nodata  & \nodata  & 682 $\pm$ 111 & 43.75 $\pm$ 0.08 & 0.0364 & Sy1.2 \\   
35 & 79 & UGC 03142 & \nodata  & \nodata  & 1581 $\pm$ 79 & 43.51 $\pm$ 0.07 & 0.0217 & Sy1 \\   
36 & 80 & 2MASX J04440903+2813003 & 90 $\pm$ 18 & \nodata  & 1427 $\pm$ 214 & 43.25 $\pm$ 0.04 & 0.0113 & Sy2 \\   
37 & 83 & CGCG 420-015 & 173 $\pm$ 7 & 471 $\pm$ 22 & 624 $\pm$ 32 & 43.85 $\pm$ 0.07 & 0.0294 & Sy2 \\   
38 & 84 & ESO 033-G~002 & 163 $\pm$ 7 & 387 $\pm$ 15 & 646 $\pm$ 120 & 43.23 $\pm$ 0.07 & 0.0181 & Sy2 \\   
39 & 85 & LEDA 097068 & 218 $\pm$ 13 & 581 $\pm$ 27 & 852 $\pm$ 53 & 44.36 $\pm$ 0.07 & 0.0577 & Sy1 \\   
40 & 87 & IRAS 05078+1626 & 166 $\pm$ 14 & 748 $\pm$ 46 & 957 $\pm$ 41 & 43.78 $\pm$ 0.04 & 0.0179 & Sy1.5 \\   
41 & 89 & Ark 120 & 252 $\pm$ 18 & 253 $\pm$ 33 & \nodata  & 44.17 $\pm$ 0.04 & 0.0323 & Sy1 \\   
42 & 90 & ESO 362-18 & 166 $\pm$ 31 & 366 $\pm$ 36 & 1277 $\pm$ 88 & 43.29 $\pm$ 0.04 & 0.0125 & Sy1.5 \\   
43 & 93 & PKS 0521-36 & 97 $\pm$ 0.1 & 216 $\pm$ 20 & \nodata  & 44.32 $\pm$ 0.06 & 0.0553 & Blazar/BL Lac \\   
44 & 109 & NGC 2110 & 300 $\pm$ 19 & 566 $\pm$ 30 & 4594 $\pm$ 47 & 43.63 $\pm$ 0.04 & 0.0078 & Sy2 \\   
45 & 110 & MCG +08-11-011 & 340 $\pm$ 18 & 1283 $\pm$ 35 & 2377 $\pm$ 62 & 43.91 $\pm$ 0.04 & 0.0205 & Sy1.5 \\   
46 & 112 & 2MASX J05580206-3820043 & 348 $\pm$ 15 & 536 $\pm$ 17 & \nodata  & 43.96 $\pm$ 0.04 & 0.0339 & Sy1 \\   
47 & 113 & IRAS 05589+2828 & 201 $\pm$ 7 & 454 $\pm$ 33 & 955 $\pm$ 79 & 44.17 $\pm$ 0.05 & 0.0330 & Sy1 \\   
48 & 114 & ESO 005-G~004 & 537 $\pm$ 46 & 520 $\pm$ 46 & 8501 $\pm$ 407 & 42.54 $\pm$ 0.05 & 0.0062 & Sy2 \\   
49 & 115 & Mrk 3 & 321 $\pm$ 5 & 1890 $\pm$ 49 & 2939 $\pm$ 282 & 43.76 $\pm$ 0.04 & 0.0135 & Sy2 \\   
50 & 118 & ESO 490-IG026 & \nodata  & \nodata  & 1409 $\pm$ 41 & 43.72 $\pm$ 0.04 & 0.0248 & Sy1.2 \\   
51 & 121 & Mrk 6 & 180 $\pm$ 8 & 522 $\pm$ 23 & 920 $\pm$ 55 & 43.73 $\pm$ 0.04 & 0.0188 & Sy1.5 \\   
52 & 122 & UGC 03601 & \nodata  & \nodata  & 385 $\pm$ 23 & 43.41 $\pm$ 0.06 & 0.0171 & Sy1.5 \\   
53 & 125 & Mrk 79 & 276 $\pm$ 6 & 611 $\pm$ 38 & 1358 $\pm$ 64 & 43.68 $\pm$ 0.04 & 0.0222 & Sy1.2 \\   
54 & 127 & Mrk 10 & \nodata  & \nodata  & 918 $\pm$ 44 & 43.73 $\pm$ 0.07 & 0.0293 & Sy1.2 \\   
55 & 131 & 2MASX J07595347+2323241 & 105 $\pm$ 22 & \nodata  & 2820 $\pm$ 120 & 43.86 $\pm$ 0.07 & 0.0292 & Sy2 \\   
56 & 132 & IC 0486 & \nodata  & \nodata  & 1213 $\pm$ 61 & 43.67 $\pm$ 0.09 & 0.0269 & Sy1 \\   
57 & 133 & Phoenix Galaxy & 274 $\pm$ 19 & 1310 $\pm$ 6 & 1196 $\pm$ 62 & 43.29 $\pm$ 0.04 & 0.0135 & Sy2 \\   
58 & 134 & FAIRALL 0272 & \nodata  & \nodata  & 791 $\pm$ 58 & 43.58 $\pm$ 0.07 & 0.0218 & Sy2 \\   
59 & 138 & FAIRALL 1146 & 157 $\pm$ 14 & 441 $\pm$ 30 & 1059 $\pm$ 152 & 43.82 $\pm$ 0.06 & 0.0316 & Sy1.5 \\   
60 & 145 & IRAS 09149-6206 & 407 $\pm$ 7 & 792 $\pm$ 17 & 1735 $\pm$ 48 & 44.33 $\pm$ 0.06 & 0.0573 & Sy1 \\
61 & 146 & Mrk 704 & 256 $\pm$ 30 & 469 $\pm$ 20 & \nodata & 43.79 $\pm$ 0.06 & 0.0292 & Sy1.5 \\   
62 & 148 & MCG -01-24-012 & \nodata  & 263 $\pm$ 44 & \nodata  & 43.55 $\pm$ 0.06 & 0.0196 & Sy2 \\   
63 & 149 & MCG +04-22-042 & 78 $\pm$ 15 & 178 $\pm$ 47 & \nodata  & 43.97 $\pm$ 0.05 & 0.0323 & Sy1.2 \\   
64 & 152 & Mrk 705 & 99 $\pm$ 21 & 213 $\pm$ 44 & 926 $\pm$ 84 & 43.56 $\pm$ 0.08 & 0.0291 & Sy1.2 \\   
65 & 153 & NGC 2992 & 299 $\pm$ 49 & 827 $\pm$ 54 & 9220 $\pm$ 194 & 42.76 $\pm$ 0.05 & 0.0077 & Sy2 \\\hline
\end{tabular}
\end{center}
\end{table}
\addtocounter{table}{-1}
\begin{table*}[!ht]
\begin{center}
\caption{{\it Continued.}}
\begin{tabular}{lllrrrccl}\hline\hline
Serial & BAT & Object name & Flux$_{09}$ & Flux$_{18}$ & Flux$_{90}$ & log $L_{\rm X}$ & Redshift & Type\\
No. & No. & & (mJy) & (mJy) & (mJy) & (erg~s$^{-1}$) &  & \\
(1) & (2) & (3) & (4) & (5) & (6) &(7) & (8) & (9)\\\hline
66 & 154 & MCG -05-23-016 & 384 $\pm$ 14 & 1391 $\pm$ 21 & 1277 $\pm$ 84 & 43.48 $\pm$ 0.04 & 0.0085 & Sy2 \\   
67 & 158 & NGC 3081 & 167 $\pm$ 13 & 699 $\pm$ 41 & 2364 $\pm$ 131 & 43.12 $\pm$ 0.04 & 0.0080 & Sy2 \\   
68 & 159 & NGC 3079 & \nodata  & 1561 $\pm$ 83 & 59311 $\pm$ 1906 & 41.97 $\pm$ 0.05 & 0.0037 & Sy2/CT \\   
69 & 163 & ESO 374-G~044 & \nodata  & 307 $\pm$ 74 & \nodata  & 43.66 $\pm$ 0.08 & 0.0284 & Sy2 \\   
70 & 164 & NGC 3227 & 444 $\pm$ 71 & 1128 $\pm$ 44 & 10596 $\pm$ 535 & 42.63 $\pm$ 0.04 & 0.0039 & Sy1.5 \\
71 & 165 & NGC 3281 & 415 $\pm$ 9 & 1509 $\pm$ 29 & 6011 $\pm$ 458 & 43.31 $\pm$ 0.04 & 0.0107 & Sy2/CT \\   
72 & 167 & LEDA 093974 & 96 $\pm$ 23 & 256 $\pm$ 68 & 941 $\pm$ 26 & 43.62 $\pm$ 0.07 & 0.0239 & Sy2 \\   
73 & 171 & NGC 3516 & 262 $\pm$ 20 & 651 $\pm$ 16 & 1317 $\pm$ 83 & 43.29 $\pm$ 0.04 & 0.0088 & Sy1.5 \\   
74 & 179 & NGC 3783 & 502 $\pm$ 10 & 1530 $\pm$ 41 & 2716 $\pm$ 108 & 43.56 $\pm$ 0.04 & 0.0097 & Sy1 \\   
75 & 185 & NGC 3998 & 98 $\pm$ 22 & 133 $\pm$ 20 & 451 $\pm$ 16 & 41.87 $\pm$ 0.06 & 0.0035 & LINER$^{\rm a}$ \\  
76 & 187 & LEDA 38038 & 166 $\pm$ 8 & 614 $\pm$ 41 & 1631 $\pm$ 123 & 43.83 $\pm$ 0.06 & 0.0280 & Sy2 \\   
77 & 188 & NGC 4051 & 346 $\pm$ 30 & 885 $\pm$ 42 & 4557 $\pm$ 254 & 41.66 $\pm$ 0.04 & 0.0023 & Sy1.5 \\   
78 & 189 & ARK 347 & \nodata  & 102 $\pm$ 56 & \nodata  & 43.59 $\pm$ 0.06 & 0.0224 & Sy2 \\   
79 & 190 & NGC 4102 & 1082 $\pm$ 49 & 3287 $\pm$ 74 & 54050 $\pm$ 2235 & 41.61 $\pm$ 0.06 & 0.0028 & LINER/CT \\   
80 & 191 & Mrk 198 & \nodata  & \nodata  & 584 $\pm$ 97 & 43.40 $\pm$ 0.08 & 0.0242 & Sy2 \\   
81 & 192 & NGC 4138 & \nodata  & \nodata  & 2161 $\pm$ 74 & 41.82 $\pm$ 0.05 & 0.0030 & Sy1.9 \\   
82 & 193 & NGC 4151 & 1032 $\pm$ 19 & 3629 $\pm$ 72 & 4594 $\pm$ 126 & 43.13 $\pm$ 0.04 & 0.0033 & Sy1.5 \\   
83 & 195 & NGC 4235 & \nodata  & \nodata  & 401 $\pm$ 50 & 42.49 $\pm$ 0.09 & 0.0080 & Sy1 \\   
84 & 196 & Mrk 766 & 220 $\pm$ 13 & 859 $\pm$ 20 & 3312 $\pm$ 190 & 42.91 $\pm$ 0.05 & 0.0129 & Sy1.5 \\   
85 & 197 & M 106 & 387 $\pm$ 47 & 449 $\pm$ 72 & \nodata  & 41.10 $\pm$ 0.09 & 0.0015 & LINER \\   
86 & 200 & NGC 4388 & 462 $\pm$ 21 & 1589 $\pm$ 31 & 10349 $\pm$ 637 & 43.69 $\pm$ 0.04 & 0.0084 & Sy2 \\   
87 & 204 & 3C 273 & 276 $\pm$ 3 & 454 $\pm$ 7 & 814 $\pm$ 56 & 46.28 $\pm$ 0.04 & 0.1583 & Blazar/FSRQ \\   
88 & 206 & NGC 4507 & 510 $\pm$ 4 & 1163 $\pm$ 31 & 4370 $\pm$ 117 & 43.80 $\pm$ 0.04 & 0.0118 & Sy2 \\   
89 & 207 & ESO 506-G027 & 114 $\pm$ 19 & 207 $\pm$ 1 & 613 $\pm$ 73 & 44.24 $\pm$ 0.04 & 0.0250 & Sy2 \\   
90 & 209 & NGC 4593 & \nodata  & 569 $\pm$ 5 & \nodata  & 43.20 $\pm$ 0.04 & 0.0090 & Sy1 \\   
91 & 211 & WKK 1263 & \nodata  & 170 $\pm$ 24 & 769 $\pm$ 29 & 43.68 $\pm$ 0.08 & 0.0244 & Sy2$^{\rm a}$ \\  
92 & 215 & 3C 279 & 265 $\pm$ 57 & 626 $\pm$ 2 & 2024 $\pm$ 117 & 46.46 $\pm$ 0.06 & 0.5362 & Blazar/FSRQ \\   
93 & 221 & NGC 4945 & 8811 $\pm$ 325 & 9945 $\pm$ 495 & \nodata  & 42.38 $\pm$ 0.04 & 0.0019 & LINER$^{\rm b}$/CT \\  
94 & 222 & ESO 323-077 & 472 $\pm$ 3 & 902 $\pm$ 17 & 6945 $\pm$ 236 & 43.33 $\pm$ 0.06 & 0.0150 & Sy1.2 \\   
95 & 223 & NGC 4992 & 59 $\pm$ 10 & \nodata  & 352 $\pm$ 42 & 43.89 $\pm$ 0.04 & 0.0251 & Sy2$^{\rm c}$ \\  
96 & 225 & MCG -03-34-064 & 453 $\pm$ 11 & 1873 $\pm$ 45 & 4634 $\pm$ 106 & 43.23 $\pm$ 0.06 & 0.0165 & Sy1.8 \\   
97 & 226 & Cen A & 10191 $\pm$ 2305 & 13148 $\pm$ 1049 & 102187 $\pm$ 12824 & 42.78 $\pm$ 0.04 & 0.0018 & Sy2/RL \\   
98 & 228 & MCG -06-30-015 & 280 $\pm$ 32 & 591 $\pm$ 11 & 1035 $\pm$ 119 & 42.97 $\pm$ 0.04 & 0.0077 & Sy1.2 \\   
99 & 229 & NGC 5252 & \nodata  & \nodata  & 415 $\pm$ 120 & 43.94 $\pm$ 0.04 & 0.0230 & Sy1.9 \\   
100 & 230 & 4U 1344-60 & 207 $\pm$ 9 & 556 $\pm$ 28 & \nodata  & 43.50 $\pm$ 0.04 & 0.0129 & Sy1.5 \\   
101 & 231 & IC 4329A & 769 $\pm$ 12 & 1790 $\pm$ 34 & 1785 $\pm$ 210 & 44.23 $\pm$ 0.04 & 0.0160 & Sy1.2 \\   
102 & 233 & Mrk 279 & 141 $\pm$ 9 & 388 $\pm$ 28 & \nodata  & 43.99 $\pm$ 0.04 & 0.0304 & Sy1.5 \\   
103 & 235 & Circinus Galaxy & 13910 $\pm$ 173 & 41447 $\pm$ 265 & 253567 $\pm$ 64752 & 42.03 $\pm$ 0.04 & 0.0014 & Sy2$^{\rm d}$/CT \\  
104 & 236 & NGC 5506 & 823 $\pm$ 27 & 2240 $\pm$ 69 & 8413 $\pm$ 302 & 43.29 $\pm$ 0.04 & 0.0062 & Sy1.9 \\   
105 & 237 & NGC 5548 & 157 $\pm$ 5 & 409 $\pm$ 40 & 1073 $\pm$ 237 & 43.68 $\pm$ 0.04 & 0.0172 & Sy1.5 \\   
106 & 238 & ESO 511-G030 & \nodata  & \nodata  & 847 $\pm$ 150 & 43.67 $\pm$ 0.06 & 0.0224 & Sy1 \\   
107 & 239 & SBS 1419+480 & 46 $\pm$ 6 & 189 $\pm$ 13 & \nodata  & 44.34 $\pm$ 0.07 & 0.0723 & Sy1.5 \\   
108 & 241 & Mrk 817 & 188 $\pm$ 10 & 669 $\pm$ 27 & 1575 $\pm$ 60 & 43.64 $\pm$ 0.07 & 0.0314 & Sy1.5 \\   
109 & 242 & NGC 5728 & 176 $\pm$ 21 & 418 $\pm$ 31 & 11383 $\pm$ 420 & 43.26 $\pm$ 0.04 & 0.0093 & Sy2/CT \\   
110 & 246 & WKK 4438 & 94 $\pm$ 15 & 266 $\pm$ 12 & 1105 $\pm$ 86 & 43.01 $\pm$ 0.10 & 0.0160 & Sy1 \\   
111 & 247 & IC 4518A & 243 $\pm$ 27 & 677 $\pm$ 22 & 7039 $\pm$ 243 & 43.24 $\pm$ 0.06 & 0.0163 & Sy2 \\   
112 & 248 & Mrk 841 & 126 $\pm$ 13 & 372 $\pm$ 25 & \nodata  & 43.89 $\pm$ 0.07 & 0.0364 & Sy1 \\   
113 & 254 & NGC 5899 & \nodata  & \nodata  & 4683 $\pm$ 184 & 42.50 $\pm$ 0.08 & 0.0086 & Sy2 \\   
114 & 257 & MCG -01-40-001 & \nodata  & \nodata  & 2304 $\pm$ 108 & 43.59 $\pm$ 0.06 & 0.0227 & Sy2 \\   
115 & 258 & Mrk 290 & \nodata  & 151 $\pm$ 13 & \nodata  & 43.65 $\pm$ 0.06 & 0.0296 & Sy1 \\   
116 & 264 & NGC 5995 & 325 $\pm$ 19 & 671 $\pm$ 4 & 4580 $\pm$ 329 & 43.76 $\pm$ 0.06 & 0.0252 & Sy2 \\   
117 & 277 & Mrk 1498 & 67 $\pm$ 12 & 214 $\pm$ 21 & \nodata  & 44.42 $\pm$ 0.04 & 0.0547 & Sy1.9 \\   
118 & 285 & 3C 345 & 102 $\pm$ 7 & 353 $\pm$ 8 & \nodata  & 46.37 $\pm$ 0.08 & 0.5928 & Blazar/FSRQ \\   
119 & 292 & NGC 6240 & 350 $\pm$ 15 & 1489 $\pm$ 12 & 23179 $\pm$ 433 & 43.94 $\pm$ 0.04 & 0.0245 & Sy2/CT \\
120 & 295 & 1RXS J165605.6-520345 & 135 $\pm$ 13 & 248 $\pm$ 24 & 1109 $\pm$ 84 & 44.35 $\pm$ 0.08 & 0.0540 & Sy1.2 \\   
121 & 317 & NGC 6300 & 277 $\pm$ 24 & 1336 $\pm$ 97 & 14928 $\pm$ 1066 & 42.45 $\pm$ 0.04 & 0.0037 & Sy2 \\   
122 & 360 & [HB89] 1821+643 & 131 $\pm$ 4 & 326 $\pm$ 10 & 635 $\pm$ 34 & 45.49 $\pm$ 0.08 & 0.2970 & Sy1/RIM \\   
123 & 370 & 3C 382 & 120 $\pm$ 12 & \nodata  & \nodata  & 44.75 $\pm$ 0.04 & 0.0579 & Sy1/RL \\   
124 & 372 & FAIRALL 0049 & 411 $\pm$ 20 & 920 $\pm$ 59 & 2619 $\pm$ 180 & 43.38 $\pm$ 0.07 & 0.0202 & Sy2 \\   
125 & 374 & ESO 103-035 & 300 $\pm$ 25 & 1446 $\pm$ 12 & 1227 $\pm$ 77 & 43.60 $\pm$ 0.04 & 0.0133 & Sy2 \\   
126 & 377 & 3C 390.3 & 90 $\pm$ 10 & 242 $\pm$ 17 & \nodata  & 44.84 $\pm$ 0.04 & 0.0561 & Sy1/RL \\   
127 & 378 & Fairall 0051 & 301 $\pm$ 7 & 697 $\pm$ 62 & 1705 $\pm$ 180 & 43.27 $\pm$ 0.06 & 0.0142 & Sy1 \\   
128 & 401 & ESO 141-G~055 & 150 $\pm$ 5 & 233 $\pm$ 38 & \nodata  & 44.14 $\pm$ 0.05 & 0.0360 & Sy1 \\   
129 & 404 & 2MASX J19301380+3410495 & 130 $\pm$ 12 & 254 $\pm$ 23 & \nodata  & 44.41 $\pm$ 0.06 & 0.0629 & Sy1 \\
130 & 405 & 2MASS J19334715+3254259 & 101 $\pm$ 10 & 269 $\pm$ 8 & \nodata  & 44.19 $\pm$ 0.08 & 0.0578 & Sy1.2 \\\hline   
\end{tabular}
\end{center}
\end{table*}
\addtocounter{table}{-1}
\begin{table*}[!ht]
\begin{center}
\caption{{\it Continued.}}
\begin{tabular}{lllrrrccl}\hline\hline
Serial & BAT & Object name & Flux$_{09}$ & Flux$_{18}$ & Flux$_{90}$ & log $L_{\rm X}$ & Redshift & Type\\
No. & No. & & (mJy) & (mJy) & (mJy) & (erg~s$^{-1}$) &  & \\
(1) & (2) & (3) & (4) & (5) & (6) &(7) & (8) & (9)\\\hline
131 & 407 & NGC 6814 & \nodata  & 258 $\pm$ 29 & 6954 $\pm$ 350 & 42.63 $\pm$ 0.04 & 0.0052 & Sy1.5 \\   
132 & 415 & Cygnus A (3C 405) & \nodata  & 418 $\pm$ 17 & 2455 $\pm$ 78 & 44.89 $\pm$ 0.04 & 0.0561 & Sy2/RL \\   
133 & 418 & NGC 6860 & 155 $\pm$ 13 & 357 $\pm$ 61 & 1369 $\pm$ 73 & 43.46 $\pm$ 0.05 & 0.0149 & Sy1 \\   
134 & 420 & MCG +04-48-002 & 388 $\pm$ 11 & 538 $\pm$ 16 & 12160 $\pm$ 289 & 43.53 $\pm$ 0.04 & 0.0139 & Sy2 \\   
135 & 425 & 4C +74.26 & 147 $\pm$ 6 & 175 $\pm$ 9 & \nodata  & 45.04 $\pm$ 0.04 & 0.1040 & Sy1/RL \\   
136 & 426 & RX J2044.0+2833 & \nodata  & 199 $\pm$ 46 & \nodata  & 44.10 $\pm$ 0.07 & 0.0500 & Sy1 \\   
137 & 427 & Mrk 509 & 247 $\pm$ 18 & 499 $\pm$ 17 & \nodata  & 44.35 $\pm$ 0.04 & 0.0344 & Sy1.2 \\   
138 & 428 & IC 5063 & \nodata  & 2246 $\pm$ 26 & 3821 $\pm$ 141 & 43.35 $\pm$ 0.04 & 0.0114 & Sy2 \\   
139 & 429 & 2MASX J21140128+8204483 & 71 $\pm$ 5 & 105 $\pm$ 32 & \nodata  & 44.77 $\pm$ 0.05 & 0.0840 & Sy1/RL \\
140 & 434 & 4C 50.55 & \nodata  & 188 $\pm$ 21 & \nodata  & 44.15 $\pm$ 0.04 & 0.0200 & Sy1$^{\rm a}$/RL \\  
141 & 435 & SWIFT J212745.58+565635.6 & 211 $\pm$ 15 & 435 $\pm$ 17 & \nodata  & 43.22 $\pm$ 0.04 & 0.0147 & Sy1 \\   
142 & 441 & 1RXS J213623.1-622400 & \nodata  & 142 $\pm$ 11 & \nodata  & 44.43 $\pm$ 0.06 & 0.0588 & Sy1 \\   
143 & 445 & Mrk 520 & 146 $\pm$ 12 & 320 $\pm$ 11 & 5040 $\pm$ 100 & 43.70 $\pm$ 0.07 & 0.0266 & Sy1.9 \\   
144 & 446 & NGC 7172 & 316 $\pm$ 17 & 424 $\pm$ 44 & 8087 $\pm$ 218 & 43.44 $\pm$ 0.04 & 0.0087 & Sy2 \\   
145 & 447 & BL Lac & 247 $\pm$ 36 & 319 $\pm$ 34 & 937 $\pm$ 122 & 44.62 $\pm$ 0.05 & 0.0686 & Blazar/BL Lac \\   
146 & 449 & NGC 7213 & 360 $\pm$ 35 & \nodata  & 2943 $\pm$ 146 & 42.59 $\pm$ 0.05 & 0.0058 & Sy1.5 \\   
147 & 451 & 3C 445 & 141 $\pm$ 13 & 358 $\pm$ 6 & \nodata  & 44.44 $\pm$ 0.05 & 0.0562 & Sy1.5/RL \\   
148 & 452 & MCG +02-57-002 & \nodata  & \nodata  & 575 $\pm$ 39 & 43.54 $\pm$ 0.09 & 0.0290 & Sy1.5 \\   
149 & 455 & NGC 7314 & \nodata  & 304 $\pm$ 58 & 4499 $\pm$ 258 & 42.33 $\pm$ 0.05 & 0.0048 & Sy1.9 \\   
150 & 456 & NGC 7319 & 88 $\pm$ 14 & 158 $\pm$ 8 & 576 $\pm$ 39 & 43.60 $\pm$ 0.05 & 0.0225 & Sy2 \\   
151 & 457 & Mrk 915 & \nodata  & 482 $\pm$ 284 & \nodata  & 43.76 $\pm$ 0.06 & 0.0241 & Sy1 \\   
152 & 460 & 3C 454.3 & \nodata  & 498 $\pm$ 6 & \nodata  & 47.37 $\pm$ 0.04 & 0.8590 & Blazar/FSRQ \\   
153 & 463 & UGC 12282 & \nodata  & \nodata  & 1284 $\pm$ 124 & 43.16 $\pm$ 0.08 & 0.0170 & Sy1.9 \\   
154 & 465 & NGC 7469 & 767 $\pm$ 17 & 2692 $\pm$ 60 & 27694 $\pm$ 1738 & 43.55 $\pm$ 0.05 & 0.0163 & Sy1.2 \\   
155 & 466 & Mrk 926 & 60 $\pm$ 3 & 214 $\pm$ 36 & 647 $\pm$ 119 & 44.65 $\pm$ 0.04 & 0.0469 & Sy1.5 \\   
156 & 467 & NGC 7582 & 1368 $\pm$ 90 & 3287 $\pm$ 199 & 60906 $\pm$ 6037 & 42.63 $\pm$ 0.04 & 0.0052 & Sy2 \\   
157 & 468 & NGC 7603 & 295 $\pm$ 11 & 321 $\pm$ 12 & 1340 $\pm$ 104 & 43.91 $\pm$ 0.04 & 0.0295 & Sy1.5 \\   
158 & 473 & LCRS B232242.2-384320 & \nodata  & \nodata  & 654 $\pm$ 80 & 43.79 $\pm$ 0.08 & 0.0359 & Sy1 \\\hline \end{tabular}
\end{center}
\tablecomments{Col.~(1): serial number.
Col.~(2): Object number in the 22-month {\it Swift}/BAT hard X-ray survey catalog \citep{2010ApJS..186..378T}.
Col.~(3): Object name.
Col.~(4)(5)(6): The flux and error in 9, 18, and 90~$\mu$m taken from {\it AKARI}/PSC in units of mJy. 
Col.~(7): The logarithmic luminosity and error in hard X-ray band (14-195~keV)
taken from \citet{2010ApJS..186..378T}, in units of erg~s$^{-1}$.  
Col.~(8): Redshift.
Col.~(9): Optical AGN type taken from \citet{2010ApJS..186..378T} 
or from other literature sources as follows:    
 }
\tablerefs{
$^{\rm a}$\citet{2010A&A...518A..10V}; 
$^{\rm b}$\citet{1996A&A...308L...1M}; 
$^{\rm c}$\citet{2007A&A...467..585B}; 
$^{\rm d}$\citet{2000MNRAS.318..173M}
}
\end{table*}

\begin{table}[!ht]
\begin{center}
\caption{Summary of cross-identifications }\label{tb:CASresult}
\begin{tabular}{llrrrrrr}\hline\hline
{\it AKARI} detector & & \multicolumn{4}{c}{Source type} & Total & Excluded source\\
& & Sy1 & Sy2 & LINER & Blazar &  &\\
(1) & (2) & (3) & (4) & (5) & (6) & (7) & (8)\\\hline
IRC 9~($\mu$m) & detection & 59 & 43 & 4 & 5 & 111 & 2\\
& non-detection & 66 & 56 & 2 & 21& 145 & 10\\\hline
IRC 18~($\mu$m) & detection & 67 & 52 & 4 & 6 & 129 & 5\\
& non-detection & 57 & 47 & 2 & 20 & 126 & 8 \\\hline
FIS 90~($\mu$m) & detection & 48 & 60 & 2 & 3 & 113 & 3\\
& non-detection & 76 & 41 & 4 & 22 & 143 & 9 \\\hline
\end{tabular}
\end{center}
\tablecomments{
Col.(1): The {\it AKARI} instruments and the center bands.
Col.(3): The number of source detection and non-detection of Seyfert 1 type AGN including Seyfert 1.2 and 1.5.  
Col.(4): The number of source detection and non-detection of Seyfert 2 type AGN including Seyfert 1.8 and 1.9.
Col.(5): The number of source detection and non-detection of LINERs.
Col.(6): The number of source detection and non-detection of Blazars.
Col.(7): The total number of source detection and non-detection, i.e., sum of columns (3) through (6).
Col.(8): The number of excluded sources, which are confused or confusing sources as defined by \citet{2010ApJS..186..378T}.
}
\end{table}


\begin{table}[!ht]
\begin{center}
\caption{Summary of the fitting results}\label{tb:LLfit}
\begin{tabular}{llccccccc}\hline\hline
{\it AKARI} detector & Subsample & {\it N} & $\rho$ &$ \rho_{\cdot z}$ & {\it a} & {\it b} & $\bar{r}$ & $\sigma_r$\\
(1) & (2) & (3) & (4) & (5) & (6)  & (7) & (8) & (9)\\
\hline
IRC 9~($\mu$m) & All & 111 & 0.85 & 0.79        & $0.22  \pm 0.05$ & $0.88 \pm 0.04$ & 0.14 &0.42\\
& All, ex CT                     & 103 & 0.88 & 0.87        & $0.13  \pm 0.04$ & $0.94 \pm 0.04$ & 0.08 & 0.32\\
& Sy1                                & 59    & 0.87 & 0.85        & $0.15  \pm 0.05$ & $0.94 \pm 0.06$ & 0.09 & 0.30\\
& Sy1, ex RL & 50 & 0.87 & 0.69        & $0.15  \pm 0.05$ & $0.99 \pm 0.07$ & 0.14 & 0.26\\
& Sy2                                & 43    & 0.58 & 0.06        & $0.23  \pm 0.08$ & $0.84 \pm 0.07$ & 0.17 & 0.53\\
& Sy2, ex CT                   & 37    & 0.70 & 0.44        & $0.11  \pm  0.07$ & $0.91 \pm 0.08$ & 0.06 & 0.35\\
& CT                                  & 8       & 0.31 & $-$0.14 & $0.81  \pm 0.28$ & $0.81 \pm 0.24$ &0.87 & 0.81\\
& RL             & 13    & 0.81 & 0.89        & $0.09 \pm 0.15$  & $0.90 \pm 0.11$ & $-$0.14 & 0.39\\
& Blazar                           & 5      & 1.00 & 0.87         & $0.11  \pm 0.23$ & $0.96 \pm 0.10$ & $-$0.002 & 0.36\\
& RL + Blazar & 18 & 0.86 & 0.88      & $0.05 \pm 0.14$ & $0.95 \pm 0.07$ & $-$0.03 & 0.37\\\hline
IRC 18~($\mu$m) & All & 129 & 0.84 & 0.77        & $0.27  \pm 0.05$ & $0.89 \pm 0.04$ & 0.19 & 0.43\\
& All, exc CT                      & 120 & 0.86 & 0.84        & $0.18  \pm 0.04$ & $0.94 \pm 0.03$ & 0.14 & 0.34\\
& Sy1                                   & 67   & 0.84 & 0.75        & $0.18  \pm 0.06$ & $0.94 \pm 0.06$ & 0.13 & 0.31\\
& Sy1, ex RL                      & 58    & 0.84 & 0.68        & $0.20 \pm 0.06$ & $0.97 \pm 0.07$ & 0.17 & 0.28\\
& Sy2                                   & 52    & 0.62 & 0.31        & $0.31  \pm 0.07$ & $0.86 \pm 0.06$ & 0.27 & 0.53\\
& Sy2, ex CT                      & 45    & 0.72 & 0.57        & $0.18  \pm 0.06$ & $0.94 \pm 0.06$ & 0.17 & 0.38\\
& CT                                     & 9      & 0.55 & $-$0.12 & $0.93  \pm 0.26$ & $0.86 \pm 0.15$ & 1.00 & 0.81 \\
& RL                & 13    & 0.90 & 0.80        & $0.07 \pm 0.18$ & $0.94 \pm 0.11$ & $-$0.02 & 0.41\\
& Blazar                              & 6      & 1.00 & 0.75        & $0.04  \pm 0.16$ & $0.98 \pm 0.07$ & $-$0.03 & 0.38\\
& RL + Blazar & 19 & 0.93 & 0.84        & $0.04 \pm 0.15$  & $0.97 \pm 0.07$ & $-$0.02 & 0.39\\\hline
FIS 90~($\mu$m) & All   & 113 & 0.57 & 0.48        & $0.21    \pm 0.06$   & $0.88 \pm 0.06$ & 0.15 & 0.58\\
& All, ex CT                          & 106 & 0.65 & 0.64        & $0.10    \pm 0.05$   & $0.94 \pm 0.06$ & $-$0.08 & 0.48\\
& Sy1                                     & 48   & 0.77 & 0.66        & $0.002 \pm 0.074$ & $0.98 \pm 0.08$ & $-$0.01 & 0.39\\
& Sy1, ex RL                         & 45   & 0.75 & 0.37        & $0.007 \pm 0.079$ & $1.01 \pm 0.11$ & 0.01  & 0.39\\
& Sy2                                     & 60   & 0.33 & 0.06        & $0.32    \pm 0.08$   & $0.91 \pm 0.06$ & 0.29 & 0.63\\
& Sy2, ex CT                        & 53   & 0.46 & 0.25        & $0.17    \pm 0.07$   & $0.94 \pm 0.07$ & 0.18 & 0.53\\
& CT                                       & 8      & 0.50 & $-$0.26 & $1.24    \pm 0.16$   & $0.65 \pm 0.05$ &1.32 & 0.69\\
& RL                                        & 8      & 0.79 & 0.83       & $0.09     \pm 0.18$  &  $0.93 \pm 0.09$ & 0.02 & 0.42\\
& Blazar                                & 3      & 1.00 & 1.00        & --                                    & --                                & $-$0.37 & 0.67\\
& RL + Blazar                      & 11    & 0.91 & 0.81        & $0.08  \pm 0.18$   & $0.89 \pm 0.11$ & $-$0.08 & 0.50\\\hline
\end{tabular}
\end{center}
\tablecomments{
Col.(1): The {\it AKARI} data used for the correlation studies.
Col.(2): Subsamples used for fitting (See Sec.~\ref{LLpot}).
Col. (3): number of sources in each subsample.
Col. (4): Spearman's Rank correlation coefficient ($\rho$).
Col. (5); partial correlation coefficient ($\rho_{\cdot z}$) estimated excluding the effect of the redshift.
Col. (6): regression slope and its 1-$\sigma$ uncertainty computed with the algorithm of OLS bisector for Equation~(\ref{eq:OLS}).
Col. (7): same as column (6) but for regression intercept and its 1-$\sigma$ uncertainty.
Col. (8) (9): average and standard deviation of the luminosity ratio $r$ = log$(L_{\rm IR}/L_{\rm X})$.
}
\end{table}%

\begin{table}[!ht]
\caption{Properties of Compton-thick AGN
\label{tb:compt}
}
\begin{center}
\begin{tabular}{ccccc}\hline\hline
No. & Object Name & Type & $N_{\rm H}$                          & Observation \\
       &                           &            & (10$^{24}$~cm$^{-2}$) &         \\
(1)   & (2) & (3) & (4) & (5) \\\hline
42    & NGC~1068         & Sy2     & $\ge$10 & {\it Beppo}SAX$^{\rm a}$ \\
60    & NGC~1365         & Sy1.8 & 4.0 & {\it Suzaku}$^{\rm b}$ \\
159 & NGC~3079         & Sy2     & 10  & {\it Beppo}SAX$^{\rm c}$ \\
165 & NGC~3281         & Sy2     & 2.0  & {\it Beppo}SAX$^{\rm d}$ \\
190 & NGC~4102         & LINER  & 2.0 & {\it Swift}/BAT$^{\rm e}$ \\
221 & NGC~4945         & LINER  & 5.3 & {\it Suzaku}$^{\rm f}$ \\
235 & Circinus Galaxy & Sy2     & 4.3 &  {\it Beppo}SAX$^{\rm g}$ \\
242 & NGC~5728         & Sy2     & 1.4 & {\it Suzaku}$^{\rm h}$ \\
292 & NGC~6240         & Sy2     & 2.2 & {\it Beppo}SAX$^{\rm i}$ \\\hline
\end{tabular}
\end{center}
\tablecomments{
Col.(1): Object number listed in the 22-month {\it Swift}/BAT hard X-ray survey catalogue \citep{2010ApJS..186..378T}.
Col.(2): Object name.
Col.(3): AGN type taken from \citet{2010ApJS..186..378T}.
Col.(4): Column density in units of 10$^{24}$~cm$^{-2}$.
Col.(5): The observation of the sources which referred following literatures.  
}
\tablerefs{
$^{\rm a}$\citet{1997A&A...325L..13M};
$^{\rm b}$\citet{2009ApJ...705L...1R};
$^{\rm c}$\citet{2001ApJ...561L..69I};
$^{\rm d}$\citet{2002A&A...381..834V};
$^{\rm e}$\citet{2011A&A...527A.142G};
$^{\rm f}$\citet{2008PASJ...60S.251I};
$^{\rm g}$\citet{1999A&A...341L..39M};
$^{\rm h}$\citet{2010ApJ...717..787C};
$^{\rm i}$\citet{1999A&A...349L..57V}
}
\end{table}

\begin{table}[!ht]
\begin{center}
\caption{The sample of starburst galaxies
\label{tb:sbg}}
\begin{tabular}{cccccc}\hline\hline
Object name  & $F_{\rm 9 \mu m}$  & $F_{\rm 90 \mu m}$  & $F_{\rm 2-10keV}$& $F_{\rm 14-195keV}$ & {\it D}  \\
& (Jy) & (Jy) & (10$^{-13}$~erg~cm$^{-2}$~s$^{-1}$) & (10$^{-13}$~erg~cm$^{-2}$~s$^{-1}$) & (Mpc) \\
(1) & (2)              & (3)                 & (4)                     & (5)                       & (6)                      \\\hline            
M82              & $70.0\pm 7.00^{\rm a}$ & $2040\pm 410^{\rm a}$ & \nodata                 & $54.0\pm 32.0^{\rm b}$ &3.50$^{\rm a}$\\
NGC~253    & $22.6 \pm 1.45$                & $1200\pm 240^{\rm c}$ & 12.3$^{\rm d}$ & 15.7 (power-law) $\pm$ 1.57           &2.58$^{\rm d}$\\
                    &                         &                           &                                                                                     & 1.76 (APEC)           $\pm$ 0.18           &\\
Arp~220     & $0.25 \pm 0.01$          & $91 \pm 63$            & 0.98$^{\rm e}$ & 1.25 (power-law)  $\pm$ 0.13           &72.9$^{\rm e}$\\
                   &                         &                                                         &                                                      &  0.13 (APEC)           $\pm$ 0.01           &\\
NGC~2146 & $5.52 \pm 0.01$          & $168 \pm 15$            & 11.4$^{\rm f}$  & 14.7 (power-law)  $\pm$ 1.47           &16.5$^{\rm j}$\\
                   &                         &                                                         &                                                       & 1.65 (APEC)           $\pm$ 0.17           &\\
NGC~3256 & $2.39 \pm 0.01$          & $108 \pm 8$            &  5.86$^{\rm g}$ & 7.59 (power-law)  $\pm$ 0.76           &35.4$^{\rm j}$\\
                   &                         &                                                         &                                                       & 0.85 (APEC)           $\pm$ 0.09           &\\
NGC~3310 & $1.08 \pm 0.02$          & $35 \pm 1$            & 21.2$^{\rm h}$  & 27.2 (power-law)  $\pm$ 2.72           &19.8$^{\rm j}$\\
                   &                         &                                                         &                                                       & 3.05 (APEC)           $\pm$ 0.31           &\\
NGC~7714 & $0.29 \pm 0.01$          & $11 \pm 1$            & 1.58$^{\rm i}$   & 2.03 (power-law)  $\pm$ 0.20           &38.2$^{\rm j}$\\
                  &                         &                                                         &                                                       &  0.23 (APEC)          $\pm$ 0.02           &\\\hline
\end{tabular}
\end{center}
\tablecomments{
Col.(1): Object name.
Col.(2)(3): Flux and error in 9 and 90~$\mu$m in units of Jy, taken from {\it AKARI}/PSC and the references below. 
Col.(4): Flux in 2-10~keV band in units of $10^{-13}$ erg~sec$^{-1}$~cm$^{-2}$, taken from the references below.
Col.(5): Flux and error in 14-195~keV band in units of 10$^{-13}$ erg~sec$^{-1}$~cm$^{-2}$. 
Except for M82, we simulated the value by using two models, power-law ($\Gamma$=2.1) and {\tt APEC} (7~keV).
Col(6): Distance in Mpc.
}
\tablerefs{
$^{\rm a}$\cite{2010A&A...514A..14K};
$^{\rm b}$\cite{2010A&A...524A..64C};
$^{\rm c}$\cite{2009ApJ...698L.125K};
$^{\rm d}$\cite{2008A&A...489.1029B};
$^{\rm e}$\cite{2009ApJ...691..261T};
$^{\rm f}$\cite{1999ApJ...514..772D};
$^{\rm g}$\cite{1999ApJ...526..649M};
$^{\rm h}$\cite{1998MNRAS.301..915Z};
$^{\rm i}$\cite{2004A&A...422..915S};
$^{\rm j}$\cite{2006ApJ...653.1129B}
}
\end{table}

\end{document}